\documentclass[aip,jcp,graphicx,reprint,longbibliography,noeprint,citeautoscript]{revtex4-1}

\usepackage[pdftex]{graphics}
\usepackage{amssymb,amsmath}
\usepackage{siunitx}

\newcommand{\tr}{\mathrm{tr}}
\newcommand{\el}{\mathrm{el}}
\newcommand{\rd}{\mathrm{d}}
\newcommand{\kBT}{\ensuremath{k_\mathrm{B}T}}
\renewcommand{\Re}{\operatorname{Re}}

\newcommand{\rev}[1]{\textcolor{black}{#1}} % First revision
\newcommand{\Rev}[1]{\textcolor{black}{#1}} % Second revision

\begin{document}

\title{Charge transport in organic semiconductors from the mapping approach to surface hopping}
\author{Johan E. Runeson}
\email{johan.runeson@chem.ox.ac.uk}
\affiliation{Department of Chemistry, University of Oxford, Physical and Theoretical Chemistry Laboratory, South Parks Road, Oxford, OX1 3QZ, UK}
\author{Thomas J. G. Drayton}
\affiliation{Department of Chemistry, University of Oxford, Physical and Theoretical Chemistry Laboratory, South Parks Road, Oxford, OX1 3QZ, UK}
\author{David E. Manolopoulos}
\affiliation{Department of Chemistry, University of Oxford, Physical and Theoretical Chemistry Laboratory, South Parks Road, Oxford, OX1 3QZ, UK}

\begin{abstract}
We describe how to simulate charge diffusion in organic semiconductors using a recently introduced mixed quantum-classical method, the mapping approach to surface hopping (MASH). In contrast to standard fewest-switches surface hopping, this method propagates the classical degrees of freedom deterministically on the most populated adiabatic electronic state. This correctly preserves the equilibrium distribution of a quantum charge coupled to classical phonons, allowing one to time-average along trajectories to improve the statistical convergence of the calculation. We illustrate the method with an application to a standard model for the charge transport in the direction of maximum mobility in crystalline rubrene. Because of its consistency with the equilibrium distribution, the present method gives a time-dependent diffusion coefficient that plateaus correctly to a long-time limiting value. \Rev{The resulting mobility is somewhat higher than that of the relaxation time approximation, which uses a phenomenological relaxation parameter to obtain a non-zero diffusion coefficient from a calculation with static phonon disorder. However, it is very similar to the mobility obtained from Ehrenfest dynamics, at least in the parameter regimes we have investigated here. This is somewhat surprising because Ehrenfest dynamics overheats the electronic subsystem and is therefore inconsistent with the equilibrium distribution}.
\end{abstract}

\maketitle

\section{Introduction}
Soft organic semiconductors have attracted considerable attention as an emerging class of materials for organic light-emitting diodes, field-effect transistors, and photovoltaic devices.\cite{Fratini2020,Ghosh2020account}
Their conductive properties can be understood in terms of charges diffusing through the material due to interaction with phonons.
For the high-mobility materials of interest, the charge-phonon coupling is too weak to be explained entirely by a local hopping (polaronic) mechanism, yet too large for an entirely delocalised (band-like) mechanism.\cite{Oberhofer2017}
In recent years, a picture has emerged of an intermediate scenario in which dynamic disorder gives rise to transient localization.\cite{Fratini2016,Nematiaram2020perspective,Giannini2023natmat}
This perspective has led to estimates of the charge mobility of a wide range of semiconductors within the relaxation time approximation.\cite{Fratini2017natmat,Nematiaram2019practical,Harrelson2019probe,Landi2019mobility} 
However, it remains unclear how well this phenomenological picture compares to direct simulation due to the lack of reliable methods to simulate the coupled dynamics of charges and phonons.

While exact quantum methods have recently been used to address this problem,\cite{DeFilippis2015qmc,Shuai2021mps,Shi2024finite,Ostmeyer2023hmc} they have so far been restricted to smaller systems and/or shorter time scales than we shall consider here.
A more tractable strategy is to use mixed quantum-classical trajectories. In a pioneering study, Troisi and Orlandi showed that even the simple Ehrenfest approximation can capture the dynamical localization of a charge coupled to intermolecular phonons.\cite{Troisi2006} However, the Ehrenfest approach is known to bring the system out of thermal equilibrium by leaking nuclear energy to the electronic subsystem.\cite{Parandekar2005mixed} As a result, the calculated diffusivity does not reach a plateau within a realistic time scale.\cite{Ciuchi2011,Fratini2016} 
Another popular quantum-classical method is Tully's fewest switches surface hopping (FSSH),\cite{Tully1990hopping} variations of which are widely used to simulate organic semiconductors.\cite{Wang2013flexible,Wang2015transport,Giannini2019,Xie2020semiconductor,Sneyd2021sciadv,Roosta2022hopping,Peng2022exciton,Giannini2022review} However, the treatment of electronic coherences in FSSH remains contentious \cite{Wang2020review}, and it is challenging to resolve the low hopping probabilities associated with a large number of weakly coupled electronic states.\cite{Wang2013flexible} \Rev{Indeed much effort has recently been devoted to detecting and correcting for missed hops in FSSH calculations.\cite{Bai2018trivial,Carof2019mobility} }

Despite their drawbacks, Ehrenfest dynamics and FSSH (with appropriate corrections) have been found to yield reasonable agreement with experiment for a variety of systems.\cite{Nematiaram2020perspective}
However, a remaining practical limitation is that the results of both methods need to be averaged over a large number of trajectories to obtain well-converged mobilities. We find it  remarkable that so little progress has been made to reduce this effort when calculating linear response properties such as the thermal mobility. For adiabatic dynamics on a single electronic potential energy surface, it has long been standard practice to accelerate the calculation of time-correlation functions by averaging along equilibrium trajectories.\cite{Tuckerman2010book} Unfortunately, it is not possible to do this in the non-adiabatic case with Ehrenfest dynamics or FSSH, since neither of these methods preserves the equilibrium distribution of the coupled charge-phonon system.

In this paper, we shall show that one can in fact simulate charge diffusion with trajectories that preserve the quantum-classical equilibrium distribution, as is required for time averaging.
Our starting point is the recently-developed `mapping approach to surface hopping' (MASH). \cite{Mannouch2023mash,Runeson2023mash} In contrast to the stochastic surface hopping of FSSH, this approach evolves trajectories deterministically on the adiabatic state with the largest population.\footnote{Note that we are referring here to the version of MASH described in Ref.~\onlinecite{Runeson2023mash}, not to the more sophisticated ``uncoupled spheres'' MASH method that was recently developed for applications to gas-phase photochemistry (Refs.~\onlinecite{Lawrence2024sizeconsistent} and \onlinecite{Lawrence2024cyclobutanone}).}  \nocite{Lawrence2024sizeconsistent}\nocite{Lawrence2024cyclobutanone} MASH is consistent with the quantum--classical Boltzmann distribution,\cite{Runeson2023mash,Amati2023thermalization} and it has previously been shown to be successful for simulating excitonic systems with as many as eight coupled electronic states.\cite{Runeson2023pccp,Lawrence2023mash} However, organic semiconductors pose an additional challenge in that they involve a band of many more (in practice, hundreds of) electronic states. Despite the apparent difficulty, we shall show that evolving on the most populated adiabatic state also works well in this context and is sufficient to overcome the overheating problem of Ehrenfest dynamics. In order to calculate the charge diffusion, we introduce a MASH estimator for the charge velocity operator and describe how to calculate its equilibrium time-correlation function. Since the resulting expression is time-translationally invariant, one can average it over time origins to accelerate the convergence with respect to the number of trajectories, which significantly reduces the cost of the calculation.

In Section~\ref{sec:theory}, we outline the problem of diffusion in organic semiconductors and describe our MASH methodology for solving it. In Section~\ref{sec:results}, we apply this methodology to a Su-Schrieffer-Heeger (SSH) model for the charge mobility in crystalline rubrene, a molecular semiconductor for which there is a large body of previous experimental and theoretical work. We discuss our results for the time-dependent charge diffusion coefficient $D(t)$, the temperature-dependent charge mobility $\mu(T)$, and the frequency-dependent optical conductivity $\sigma(\omega)$ in light of this previous work. In particular, we highlight the differences between the present MASH results and those of Ehrenfest dynamics, the classical path approximation (CPA),\cite{Wang2011charge,Fetherolf2020prx,Fetherolf2023prb} and the relaxation time approximation (RTA),\cite{Ciuchi2011,Cataudella2011prb,Fratini2016} all three of which are still widely used to study charge transport in organic semiconductors. Our conclusions are drawn in Section~\ref{sec:conclusions}.

\section{Theory}\label{sec:theory}

\subsection{Hamiltonian and Observables}
A standard model of a charge (an electron or a hole) interacting with phonons in an organic semiconductor is provided by the SSH Hamiltonian
\begin{align}
\hat{H}(p,q) = &\sum_j \Bigl(-J + \alpha[\hat{q}_j-\hat{q}_{j+1}]\Bigr)\Bigl(|j\rangle\langle j+1| + |j+1\rangle\langle j|\Bigr) \nonumber \\
&+\sum_j \Bigl(\frac{\hat{p}_j^2}{2m}+\frac{1}{2}m\omega_0^2\hat{q}_{j}^2\Bigr), \label{H}
\end{align}
where the index $j$ runs over the molecules along a particular lattice direction in the molecular crystal. The ket $|j\rangle$ is a basis state in which the charge is located on molecule $j$. The Hamiltonian involves a nearest neighbour interaction that fluctuates around an average value of $J$ as a result of Peierls coupling to harmonic phonons. The position of the charge in the model is represented by the operator
\begin{equation}
\hat{x} = \sum_j aj\,|j\rangle\langle j|,\label{x}
\end{equation} 
where $a$ is the equilibrium intermolecular spacing between the molecules. This model assumes that the material is sufficiently anisotropic to consider only one-dimensional diffusion, and that the effect of any Holstein coupling to high frequency intramolecular phonons has already been folded into the definition of the parameters $J$ and $\alpha$ with a polaron transformation.\cite{Nematiaram2020perspective}

The key experimental observable is the mobility $\mu$ of the charge, which is related to its diffusion coefficient by the Einstein--Smoluchowski equation $\mu= {eD/k_{\rm B}T}$. The diffusion coefficient $D$ can be written as the long-time limit of a time-dependent diffusion coefficient
\begin{equation}
D(t) = \frac{1}{2}\frac{\rm d}{\rd t}\Delta x^2(t) \equiv \int_0^{t} C(t')\,{\rm d}t',\label{Dt}
\end{equation}
where $\Delta x^2(t)$ is the mean squared displacement of the charge in time $t$,
\begin{equation}
\Delta x^2(t) = \left<\left[\hat{x}(t)-\hat{x}(0)\right]^2\right>,\label{Dx2t}
\end{equation}
and $C(t)$ is its velocity autocorrelation function
\begin{equation}
C(t) = \frac{1}{2}\Bigl<\hat{v}(0)\hat{v}(t)+\hat{v}(t)\hat{v}(0) \Bigr>.\label{Ct}
\end{equation}
Here $\hat{x}(t)$ and $\hat{v}(t)$ are the Heisenberg operators $\hat{A}(t) = e^{+i\hat{H}t/\hbar}\hat{A}\,e^{-i\hat{H}t/\hbar}$, the velocity operator $\hat{v}=(i/\hbar)[\hat{H},\hat{x}]$ is the Heisenberg time derivative of $\hat{x}$, and the expectation values are with respect to the thermal equilibrium distribution
\begin{equation}
\left<\cdots\right> = {{\displaystyle {\rm tr}\left[e^{-\beta \hat{H}}(\cdots)\right]}\over 
{\displaystyle {\rm tr}\left[e^{-\beta \hat{H}}\right]}},\label{QMaverage}
\end{equation}
where $\beta=1/k_{\rm B}T$. 

More detailed information about the diffusive process can be obtained by measuring the frequency-dependent optical conductivity $\sigma(\omega)$. This is related to $C(t)$ by\cite{Ciuchi2011}
\begin{equation}
\sigma(\omega) = ne^2\,{\tanh(\beta\hbar\omega/2)\over\hbar\omega/2}\,{\rm Re} \int_0^{\infty} e^{+i\omega t}C(t)\,{\rm d}t,
\end{equation}
where $n$ is the number density of charges in the crystal. The low-frequency limit of $\sigma(\omega)$ is directly related to $\mu$ through $\sigma(\omega\to 0)=ne\mu$, and its behaviour at higher frequencies is readily available from a calculation of the velocity autocorrelation function $C(t)$.

In what follows we will assume that the temperature is sufficiently high ($k_{\rm B}T>\hbar\omega_0$) that the vibrations can be treated classically ($\hat{p}_{j}\to p_{j}$ and $\hat{q}_{j}\to q_{j}$), as is certainly the case for the rubrene model considered in Sec.~III. To avoid boundary effects in a finite chain of $N$ molecules, we will apply periodic boundary conditions by setting $|N+1\rangle\equiv |1\rangle$, $p_{N+1,k}\equiv p_{1}$, and $q_{N+1}\equiv q_{1}$ in both the Hamiltonian and the velocity operator
\begin{equation}
\hat{v}(q) = \frac{i}{\hbar}\sum_{j=1}^N a J_j(q)\Bigl(|j\rangle\langle j+1|-|j+1\rangle\langle j|\Bigr) \label{vop}
\end{equation}
where
\begin{equation}
J_j(q)= -J+\alpha (q_{j}-q_{j+1}).\label{tauj}
\end{equation}
Note in passing that the matrix representation of this velocity operator is purely imaginary in the site basis, with zero elements on the diagonal. These properties also hold in any other real basis, including the adiabatic basis introduced below. 

\subsection{Quantum-Classical Dynamics}

When periodic boundary conditions are applied to an $N$-molecule chain with classical vibrational modes as we have described above, the Hamiltonian in Eq.~\eqref{H} can be written more compactly as
\begin{equation}
\hat{H} = T(p)+\hat{V}(q),\label{HMASH}
\end{equation}
where $q=\left\{q_{j}\right\}$ and $p=\left\{p_{j}\right\}$ are the coordinates and momenta of the classical vibrations, $T(p)$ is the vibrational kinetic energy
\begin{equation}
T(p) = \sum_{j=1}^N {p_{j}^2\over 2m},\label{Tp}
\end{equation}
and $\hat{V}(q)$ is a potential energy operator 
\begin{equation}
\hat{V}(q) = \sum_{i=1}^N\sum_{j=1}^N |i\rangle V_{ij}(q) \langle j|. \label{Vd}
\end{equation}
The diagonal representation of this operator is
\begin{equation}
\hat{V}(q) = \sum_{a=1}^N |a(q)\rangle V_a(q) \langle a(q)|, \label{Va}
\end{equation}
where $V_a(q)$ are the adiabatic potential energy surfaces and $|a(q)\rangle$ are the adiabatic eigenstates at the configuration $q$.  A normalized wavefunction $|c\rangle$ for the charge can be written equivalently in either representation as
\begin{equation}
|c\rangle = \sum_{j=1}^N |j\rangle\,c_j = \sum_{a=1}^N |a(q)\rangle\,c_a.\label{|c>}
\end{equation} 

In all of the trajectory methods described below, this wavefunction satisfies the Schr\"{o}dinger equation of motion
\begin{equation}\label{cdot}
|\dot{c}\rangle = -\frac{i}{\hbar}\hat{V}(q)|c\rangle.
\end{equation}
The classical degrees of freedom satisfy equations of motion of the form
\begin{align}
\dot{q}_j &= p_j/m \label{qdot} \\ 
\dot{p}_j &= F_j(q), \label{pdot}
\end{align}
but with different definitions of the force $F_j(q)$:

\begin{itemize}
\item[(i)] In the classical path approximation (CPA), one completely neglects the feedback of the quantum state on the classical degrees of freedom. For the SSH Hamiltonian, this prescription corresponds to
\begin{equation}
F_j(q) = -m\omega_0^2 q_j.
\end{equation}
This approximation is equivalent to the approach considered in Refs.~\onlinecite{Wang2011charge,Fetherolf2020prx,Fetherolf2023prb}. When the nuclear variables are sampled from the thermal distribution of the uncoupled bath, it is also formally equivalent to solving a stochastic Schr\"{o}dinger equation.\cite{Zhong2011sse}

\item[(ii)] In (mean-field) Ehrenfest dynamics, the effective force is the expectation value of the force operator,
\begin{equation}
F_j(q) = -\langle c|\nabla_j \hat{V}(q)|c\rangle,
\end{equation}
which can be evaluated using either of the representations in Eq.~\eqref{|c>}. This method has been widely used to study charge transport in organic semiconductors.\cite{Troisi2006,Troisi2007ehrenfest,Wang2011charge,Poole2016mobility,Hegger2020jctc,Berencei2022}

\item[(iii)] In MASH, we use the force of the most populated adiabatic state. By introducing the ``step function''
\begin{equation}
\Theta_a(c) = \begin{cases} 1, & \hbox{if } |c_a|^2>|c_b|^2 \hbox{ for all }b\not=a, \\
0, & \hbox{otherwise}, \end{cases}
\end{equation}
this force can be written as
\begin{equation}\label{Fmash}
F_j(q) = - \sum_a \nabla_j V_a(q) \Theta_a(c).
\end{equation}
\end{itemize}

Since we are interested in the dynamical properties of a mixed quantum-classical system at thermal equilibrium, the relevant expectation values are with respect to quantum-classical limit of Eq.~\eqref{QMaverage}, 
\begin{equation}
\langle \cdots \rangle = \frac{\int \rd p\, \rd q\, \tr_\el[ e^{-\beta\hat{H}(p,q)}  (\cdots)]}{\int \rd p \,\rd q\, \tr_\el[ e^{-\beta\hat{H}(p,q)}]}.
\end{equation}
However, neither the CPA nor Ehrenfest dynamics is consistent with this distribution.\cite{parandekar2006ehrenfest} With these methods, even a system initially in equilibrium will experience an unphysical flow of energy from the classical to the quantum subsystem, leading to long-time populations that correspond to an elevated temperature (in the case of the CPA, this temperature is known to be infinite\cite{runeson2022fmo}). MASH, on the other hand, is consistent with the correct distribution by virtue of the identity\cite{Runeson2023mash}
\begin{equation}
\langle \cdots \rangle =  \int \rd p\, \rd q \int_{|c|=1} \rd c \, \rho(p,q,c) \sum_a \langle a(q)|\cdots|a(q)\rangle\Theta_a(c),
\label{QCaverage}
\end{equation}
where 
\begin{equation}
\rho(p,q,c) = \frac{e^{-\beta E(p,q,c)}}{\int \rd p\, \rd q \int_{|c|=1} \rd c \, e^{-\beta E(p,q,c)}}\label{rhopqc}
\end{equation}
is the Boltzmann density corresponding to the energy
\begin{equation}
E(p,q,c) = T(p)+\sum_a V_a(q)\Theta_a(c)\label{EMASH}.
\end{equation}
In these equations, $c$ plays the role of a complex vector of phase-space variables for the quantum degree of freedom rather than a wavefunction. \rev{In practice, it is most convenient to evolve $c$ in the diabatic representation to avoid dealing with local non-adiabatic couplings. At the end of each time step, we calculate the adiabatic populations and attempt a hop whenever a new state has reached a higher population than the currently active adiabat.}

To handle hops in MASH, it is helpful to think of the term $\sum_a V_a(q)\Theta_a(c)$ in Eq.~\eqref{EMASH} as an effective step potential in the phase space $(p,q,c)$. When the active adiabatic state changes from $a$ to $b$, the adiabatic potential changes from $V_a(q)$ to $V_b(q)$. If the component $p_{\perp}$ of the momentum perpendicular to the step has sufficient kinetic energy to overcome it, the transition is successful and $p_{\perp}$ is scaled in such a way as to conserve $E(p,q,c)$. If not, the transition is unsuccessful and the trajectory is reflected from the step with a reversal of $p_{\perp}$, which has the effect of restoring $a$ as the active state and enabling the nuclear motion to continue on $V_a(q)$. By analysing the terms %of the type $V_a\nabla_j \Theta_a(c)$ 
in Eq.~\eqref{Fmash}, one can show that $p_{\perp}$ is the projection of $p$ along the direction of the vector with elements\cite{Runeson2023mash}
\begin{equation}
\delta_j = \Re \sum_{a'} c_{a'}^*(d^j_{a'a}c_a - d^j_{a'b}c_b)
\end{equation}
where $d_{ab}^j=\langle a(q)|\nabla_j|b(q)\rangle$ is an element of the non-adiabatic coupling vector between adiabatic states $a$ and $b$.

This deterministic surface hopping algorithm clearly conserves both  $E(p,q,c)$ and $\rho(p,q,c)$, and since it is equivalent to integrating Eqs.~\eqref{cdot}--\eqref{pdot} across a smoothed step it also conserves the phase space volume element ${\rm d}p\,{\rm d}q\,{\rm d}c$. Hence, the MASH dynamics relaxes to the correct quantum-classical equilibrium state populations.\cite{Amati2023thermalization} This is in contrast to the stochastic FSSH algorithm, which does not in general relax to give the correct populations.\cite{schmidt2008SH} 

\subsection{Velocity autocorrelation function}\label{sec:corrfunc}

Next, we describe how to calculate the velocity autocorrelation function in Eq.~\eqref{Ct} in the different methods.
In the CPA and Ehrenfest dynamics, we first rewrite $C(t)$ as
\begin{equation}
C(t) = \frac{\int \rd p\, \rd q \sum_a e^{-\beta [T(p)+V_a(q)]} C_a(t)}{\int \rd p\, \rd q \sum_a e^{-\beta [T(p)+V_a(q)]}}
\end{equation}
where
\begin{equation}
C_a(t) = \Re \,\langle a| \hat{v}e^{+i\hat{H}t/\hbar}\hat{v}e^{-i\hat{H}t/\hbar}|a\rangle.
\end{equation}
We initialize trajectories by sampling $p$ and $q$ from the Boltzmann distribution for uncoupled phonons. Then, for each $q$, we sample an adiabatic state $a$ from the Boltzmann distribution $e^{-\beta V_a(q)}/\sum_{a} e^{-\beta V_a(q)}$ and initialize the wavefunction as $|a\rangle$. Along the trajectory, we propagate $|a(t)\rangle$ from $|a\rangle$ and $|va(t)\rangle$ from $|va\rangle\equiv \hat{v}|a\rangle$ with the Schr\"{o}dinger equation. Then we evaluate
\begin{equation}
C_a(t) = \Re\, \langle va(t)|\hat{v}| a(t)\rangle
\end{equation}
and average over trajectories. Note that this two-step sampling procedure generates an initial distribution that is not strictly the same as the true coupled Boltzmann distribution. We use it here for consistency with previous work\cite{Ciuchi2011,Fratini2016}, noting that the initial conditions should not affect long-time properties such as the mobility.

In MASH, it is possible to construct another approximation to $C(t)$ that looks more like a classical correlation function. The derivation is given in the Appendix so here we shall simply state the result: $C(t)$ can be calculated as the canonical phase space average
\begin{equation}
C(t) = \langle v(q_0,c_0)v(q_t,c_t)\rangle, \label{CMASH}
\end{equation}
where 
\begin{equation}
\langle\cdots\rangle = \int {\rm d}p_0{\rm d}q_0\int_{|c_0|=1} {\rm d}c_0\, \rho(p_0,q_0,c_0) (\cdots) \label{MASHaverage}
\end{equation}
and
\begin{equation}
v(q,c) = \sqrt{1\over 2\Gamma}\sum_{a=1}^N\sum_{b=1}^N v_{ab}(q)\Theta_{ab}(c)c_a^*c_b \label{vqc}
\end{equation}
with $v_{ab}(q)=\langle a(q)|\hat{v}(q)|b(q)\rangle$ and $\Theta_{ab}(c)=\Theta_a(c)+\Theta_b(c)$. The constant $\Gamma$ is
\begin{equation}
\Gamma = {H_N\over N(N-1)}-{H_N^2+G_N\over (N+1)N(N-1)},
\label{gammaN}
\end{equation}
where $H_N=\sum_{n=1}^N 1/n$ and $G_N = \sum_{n=1}^N 1/n^2$. 

Since the MASH dynamics conserves both the Boltzmann factor $\rho(p_0,q_0,c_0)$ and the phase-space volume element ${\rm d}p_0\,{\rm d}q_0\,{\rm d}c_0$, we can also write
\begin{equation}
C(t) = \langle v(q_{\tau},c_{\tau})v(q_{\tau+t},c_{\tau+t})\rangle \label{Cvvtau}
\end{equation}
for any time origin $\tau$. This is especially useful because it allows us to time-average along our trajectories and write the correlation function as
\begin{equation}
C(t) = {1\over T} \int_0^T {\rm d}\tau\,\langle v(q_{\tau},c_{\tau})v(q_{\tau+t},c_{\tau+t})\rangle,
\end{equation}
which significantly reduces the statistical error in the calculation. To generate an equilibrium ensemble in practice, we first pre-equilibrate by running MASH from an adiabatically sampled initial condition in the presence of a thermostat. In the calculations presented below, we found that the equilibration phase only needed to be a small fraction (10\%) of the total time used to calculate the correlation function.

\section{Results and Discussion} \label{sec:results}

\begin{figure}
    \centering
    \includegraphics{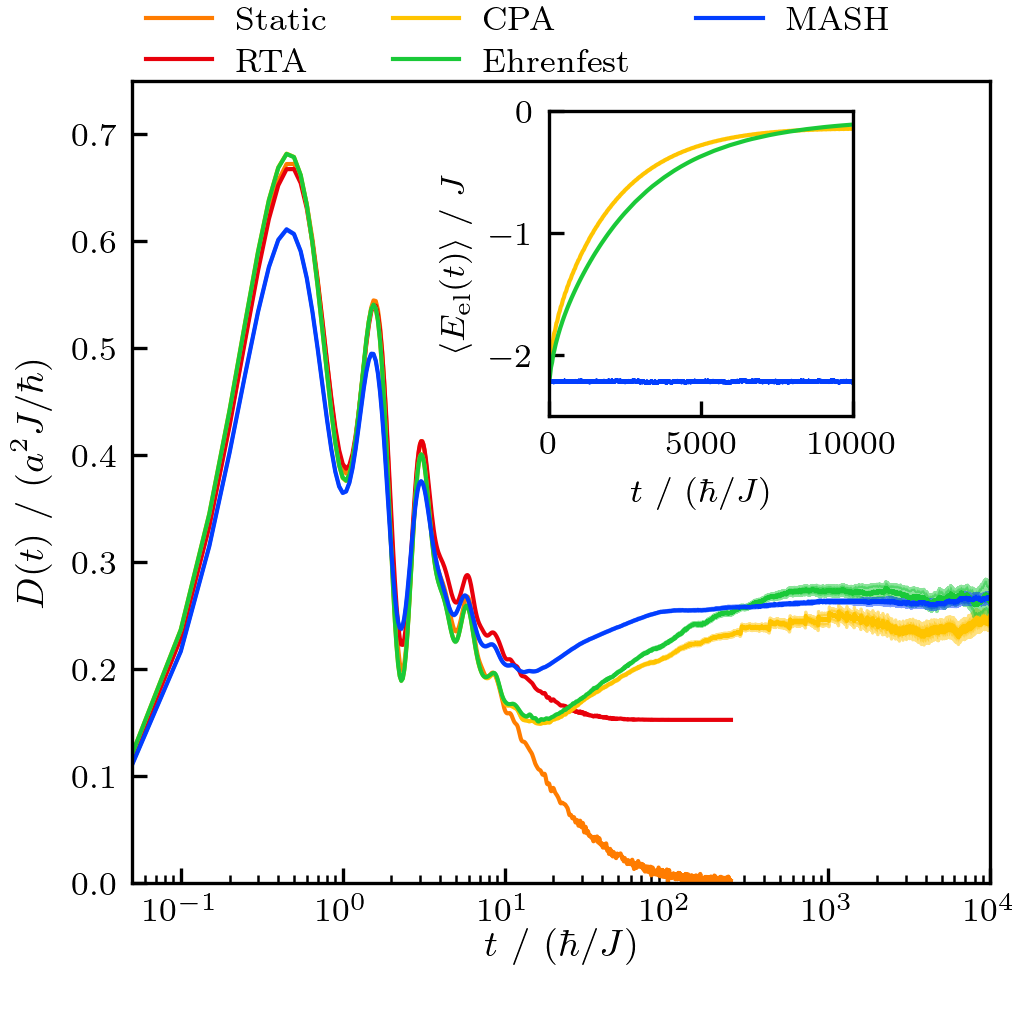}
    \caption{\rev{Time-dependent diffusivities at $T=\SI{300}{K}$ ($k_{\rm B} T=0.235\,J$) for the rubrene model. The inset shows the average electronic energy as a function of time along the Ehrenfest, CPA (no backaction), and MASH trajectories. MASH starts from an equilibrium initial condition and remains in thermal equilibrium throughout. The Ehrenfest and CPA calculations start from (a close approximation to) thermal equilibrium and eventually reach a non-equilibrium steady state in which the electronic subsystem is at an elevated temperature. The asymptotic mobilities obtained from the three calculations are very similar nevertheless.}}
    \label{fig:Dt}
\end{figure}

We have performed our simulations for the SSH model in Eq.~\eqref{H} with the parameters $J=\SI{110}{meV}$ ($\SI{887}{cm^{-1}}$), $\hbar\omega_0=0.0435J$ ($\SI{40}{cm^{-1}}$), and $\lambda=\alpha^2/(2m\omega_0^2J)=0.25$. There are several sets of slightly different parameters in the literature and these particular parameters were chosen to be consistent with Ciuchi \emph{et al.} (Ref.~\onlinecite{Ciuchi2011}). There is no need to specify the mass $m$ because it does not affect the diffusion of the charge. In practice we work with mass-scaled coordinates so $m$ does not even enter our calculation. While $N=100$ sites were sufficient to converge our MASH calculations (and also the Static and RTA calculations described below), we needed $N=200$ sites to converge our Ehrenfest and CPA calculations, which we found to be more sensitive to finite size effects (see the Supplementary Material). We used a lattice spacing of $a=\SI{7.2}{\AA}$ and a time step of $0.05\,\hbar/J$ ($\SI{0.3}{fs}$) in all of our simulations.

\subsection{Time-dependent diffusivity}

To illustrate the importance of dynamical disorder, we first consider the \emph{static} limit in which the phonon degrees of freedom are disordered at time $t=0$ and then kept frozen. For this scenario, the velocity autocorrelation function $C_0(t)$ can be easily computed in the eigenbasis of the disordered Hamiltonian, and then averaged over realisations of the disorder. As was shown by Ciuchi \emph{et al.},\cite{Ciuchi2011} and is shown again here in Fig.~\ref{fig:Dt}, the static limit leads to a time-dependent diffusivity that initially rises ballistically (super-diffusion), then falls (sub-diffusion) and eventually decays to zero as a result of Anderson localization. In other words, the mobility vanishes in this one-dimensional model when the phonons are frozen because the charge is stuck in a localised eigenstate. 

\rev{A crude estimate of the mobility can nevertheless be extracted from the static phonon calculation by invoking a relaxation time approximation (RTA),\cite{Ciuchi2011} as is done in transient localization theory.\cite{Fratini2016} The underlying assumption is that the coherent evolution in the static picture decays over some phenomenological timescale $\tau$ such that
\begin{equation}
C_\mathrm{RTA}(t) = C_0(t)e^{-t/\tau}.\label{CRTA}
\end{equation}
This is formally equivalent to applying a Lorentzian broadening to $\sigma(\omega)$,\cite{Cataudella2011prb} and it results in a finite mobility because the time integral of $C_\mathrm{RTA}(t)$ (or equivalently the zero-frequency value of $\sigma(\omega)$) is finite. Usually one chooses the relaxation time to be $\tau=\omega_0^{-1}$, i.e., to correspond to the timescale of the phonon motion ($\tau=23\,\hbar/J$ for the present model). The result of a RTA calculation with this value of $\tau$ is also shown in Fig.~\ref{fig:Dt}.}

Next, we consider what the various quantum-classical simulation methods predict about the effect of dynamical disorder. In Ehrenfest dynamics, the diffusivity follows the static curve up to $t\approx 10\,\hbar/J$, after which it increases and eventually plateaus. (Ciuchi {\em et al.}\cite{Ciuchi2011} stopped their Ehrenfest calculation with the present parameters at $t\sim 10^2 \,\hbar/J$, which gave them the impression that a plateau would not be reached.) However, because Ehrenfest dynamics suffers from an
unphysical energy leakage from the phonons to the electronic degree of freedom, the plateau does not correspond to a system at thermal equilibrium. 

\rev{The energy leakage can be monitored by calculating the average energy in the electronic subsystem along the trajectories of the Ehrenfest simulation. A suitable definition is \begin{equation}
    E_{\rm el}(q,c) = \langle c|\hat{V}_{\rm el}(q)|c\rangle,\label{EelCPA}
\end{equation}
where $\hat{V}_{\rm el}(q)$ is the operator in the first line of Eq.~\eqref{H}. The average value of this quantity along the Ehrenfest trajectories is shown in the inset of Fig.~\ref{fig:Dt}. The electronic system is clearly not in thermal equilibrium with the phonons in the plateau region, but rather in a non-equilibrium steady state in which the phonon kinetic energy is close to $Nk_{\rm B}T/2$ (for sufficiently large $N$) and yet the electronic subsystem is approaching infinite temperature ($\langle c|\hat{V}_{\rm el}(q)|c\rangle \simeq 0$).}

\rev{In the case of the CPA, the phonons are uncoupled harmonic oscillators so their average kinetic energy does not leak into the electronic subsystem -- it simply fluctuates around the thermal equilibrium value of $Nk_{\rm B}T/2$. However, the electronic subsystem still heats up because it is being driven by a periodically oscillating field, and the net effect of this on the diffusivity is much the same as in Ehrenfest dynamics.\cite{Wang2011charge} On a linear scale it may appear as though the diffusivity has reached a plateau in both of these methods by the time $t\approx 20\, \hbar/J$, at which point the overheating of the electronic subsystem has barely begun. However, it is clear from Fig.~\ref{fig:Dt} that this is not the case, so when mobilities are reported using these methods one should be made aware that they depend on the time at which the calculation was truncated.}

\rev{In the MASH calculation, the diffusivity reaches a well-defined plateau on the same time scale as the RTA. The long-time limit of the diffusivity is unambiguous, and there is no need to choose a phenomenological relaxation time. This is because MASH is consistent with the quantum-classical equilibrium distribution, which precludes any unphysical overheating of the electronic subsystem. A suitable definition of the electronic energy in MASH is the electronic part of the conserved energy $E(p,q,c)$ in Eq.~\eqref{EMASH}. This can be written as
\begin{equation}
    E_{\rm el}(q,c) = \sum_a V_{{\rm el},a}(q)\Theta_a(c),\label{EelMASH}
\end{equation}
where $V_{{\rm el},a}(q)$ are the eigenvalues of $\hat{V}_{\rm el}(q)$. As shown in the inset of Fig.~\ref{fig:Dt}, the average of this energy over the MASH trajectories fluctuates around its initial thermal equilibrium value for all time. The short-time behaviour of the diffusivity obtained from MASH does differ slightly from that of the other approaches, but that is simply because the MASH calculation starts from the coupled equilibrium distribution (as explained in Sec.~\ref{sec:corrfunc}).}

The time-averaging along the equilibrium trajectories in MASH also helps to improve the convergence of the calculation. The shaded areas in Fig.~\ref{fig:Dt} show the standard errors in the mean from 10 batches of $1000$ trajectories for MASH, and for 10 batches of $100\,000$ trajectories for Ehrenfest dynamics and the CPA. (The Static curve is shown without error bars and was obtained from a single run of 10\,000 trajectories. The RTA curve was obtained by post-processing the results of the Static calculation.) The statistical error in the MASH calculation with time averaging is clearly far smaller than that in the Ehrenfest calculation without time averaging, especially in the diffusive regime.

Unfortunately, this does not necessarily make MASH cheaper than Ehrenfest dynamics, because MASH requires an $O(N^3)$ matrix diagonalisation at each time step to find the adiabatic states whereas Ehrenfest dynamics only requires $O(N)$ operations per time step (for the present model problem) when the wavefunction is evolved in the site basis. It may be possible to find a way to avoid a full matrix diagonalisation at each time step and make the MASH calculation less expensive, which would be useful for future applications of the method to more sophisticated (2D and 3D) models of charge transport in crystalline materials. 

\begin{figure}
    \centering
    \includegraphics{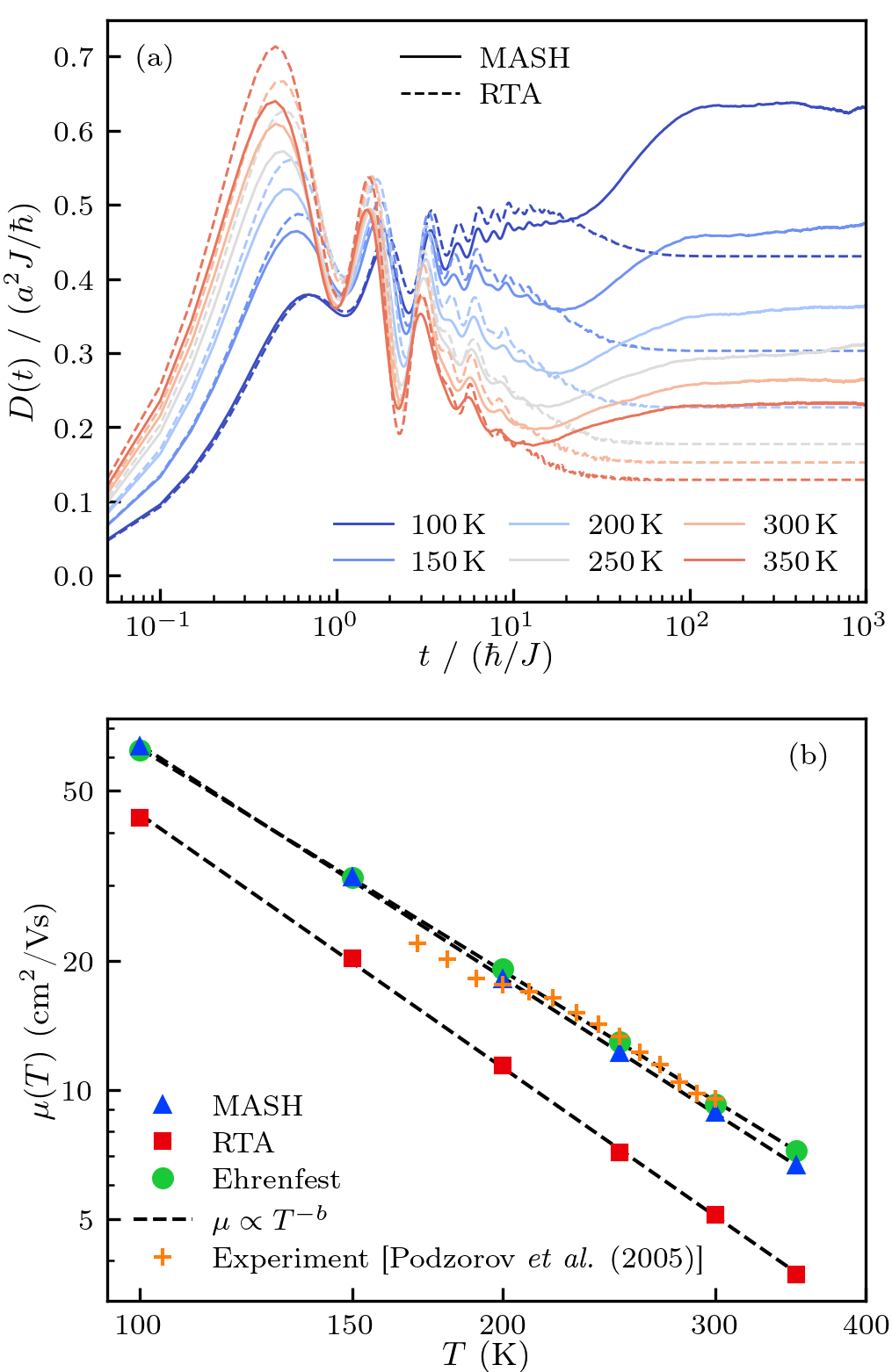}
    \caption{(a) MASH and RTA time-dependent diffusivities over a range of temperatures. (b) Log-log plot of the resulting mobilities as a function of temperature, and their fits to an inverse power law. \rev{Also shown for comparison are the results from Ehrenfest dynamics.} All three calculations are seen to be qualitatively consistent with the results of a representative experiment from Ref.~\onlinecite{Podzorov2005}.}
    \label{fig:muT}
\end{figure}

\subsection{Temperature-dependent mobility}

%We will now confine our attention to MASH and the RTA since these are the two methods that give a well-defined mobility within a reasonable simulation time ($t\sim 10^2\,\hbar/J$). 
Figure~\ref{fig:muT}(a) shows the time-dependent diffusivities obtained from the MASH and RTA methods over a range of temperatures. Each MASH (RTA) curve was calculated from a single batch of 1000 (10\,000) trajectories, and we again used a relaxation time of $\tau = \omega_0^{-1}$ in the RTA. A notable feature of the results is that the long-time plateau value of the diffusivity is consistently higher (by between 30 and 60\%) in the  MASH calculation than in the RTA. This discrepancy is comparable to the sensitivity of the RTA mobility to different choices of $\tau$.\cite{Cataudella2011prb} 

\rev{Figure~\ref{fig:muT}(b) shows a log-log plot of the mobilities $\mu = eD(\infty)/k_{\rm B}T$ obtained from the MASH and RTA calculations as a function of temperature. Also shown for comparison are the mobilities from Ehrenfest dynamics, which are found to be close to MASH across the entire temperature range. All methods give mobilities that follow an inverse power law $\mu \propto T^{-b}$, with $b=1.8$ for MASH, $b=2.0$ for the RTA (with $\tau=\omega_0^{-1}$), and $b=1.7$ for Ehrenfest dynamics. These inverse power laws are in qualitative agreement with experimental measurements as can be seen from the comparison with the mobilities of Podzorov {\em et al.}\cite{Podzorov2005} in the figure. Quantitative comparisons should be avoided, however, because the calculations are sensitive to the SSH model parameters and the experimental results are sensitive to both the quality of the crystal sample and the details of the experimental setup. The quantitative agreement between the MASH/Ehrenfest results and the experimental results in Fig.~\ref{fig:muT} is therefore fortuitous.}

\begin{figure}
    \centering
    \includegraphics{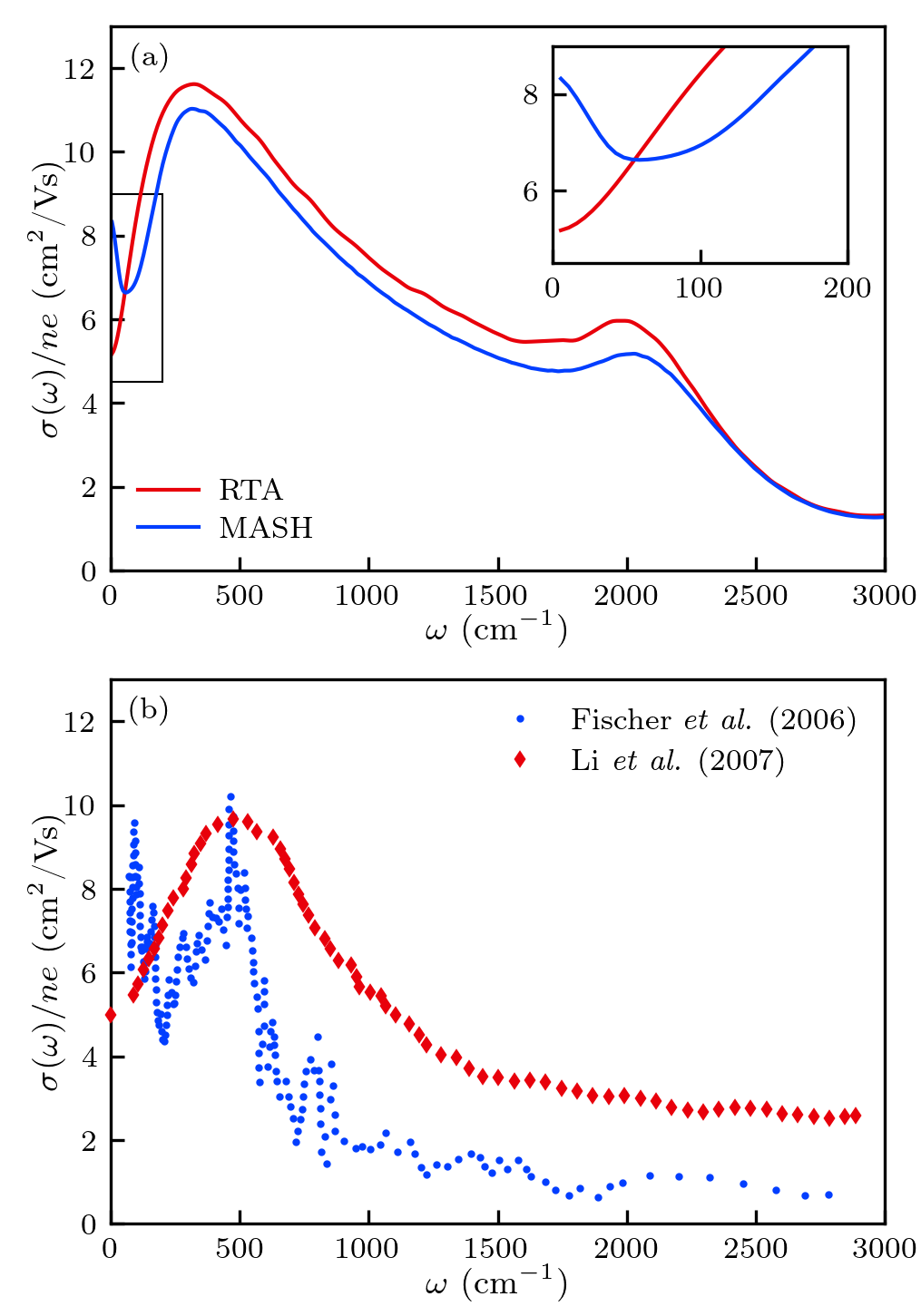}
    \caption{(a) MASH and RTA optical conductivities at $T=\SI{300}{K}$. As the frequency goes to zero (magnified in the inset), the RTA curve decreases smoothly to the zero-frequency mobility, whereas the MASH curve turns up as a result of the long-time rise of $D(t)$ observed in Fig.~\ref{fig:Dt}. (b) Available experimental results\cite{Fischer2006infrared,LiPodzorov2007} are qualitatively consistent with either of these behaviours. The experimental data have been scaled such that the lowest-frequency point is the reported DC mobility, $\SI{8.3}{cm^2/Vs}$ for Fischer \emph{et al.} (Ref.~\onlinecite{Fischer2006infrared}) and $\SI{5}{cm^2/Vs}$ for Li \emph{et al.} (Ref.~\onlinecite{LiPodzorov2007}).}
    \label{fig:sigmaw}
\end{figure}

\subsection{Frequency-dependent optical conductivity}

Further insight into the differences between the MASH and RTA calculations can be obtained by comparing their predictions for the optical conductivity. This comparison is shown for a temperature of 300 K in the upper panel of Fig.~\ref{fig:sigmaw}. Both calculations predict a main peak in $\sigma(\omega)$ at around 400 cm$^{-1}$, and a smaller peak at around 2000 cm$^{-1}$. These peaks come from short-time oscillations in the velocity autocorrelation function that give rise to the structure below $t\sim 10\,\hbar/J$ in $D(t)$, which is similar in the two calculations  (see Fig.~\ref{fig:Dt}). Where the calculations differ is in the behaviour of the optical conductivity as $\omega\to 0$. The RTA predicts a smooth decrease in $\sigma(\omega)$ below the peak at 400 cm$^{-1}$,\cite{Fratini2014phenomenological} whereas MASH predicts a low frequency rise. This rise comes from a long-time tail in the MASH velocity autocorrelation function that is responsible for the increase of the MASH $D(t)$ to its plateau value in Fig.~\ref{fig:Dt}. The tail is absent by construction in the RTA because it is not present in the static disorder calculation that produces $C_0(t)$, and even if it were present it would be eliminated by the factor of $e^{-t/\tau}$ in Eq.~\eqref{CRTA}.

The lower panel of Fig.~\ref{fig:sigmaw} shows two representative optical conductivity spectra of crystalline rubrene at 300 K. These were measured using different crystal samples in different laboratories with different experimental setups.\cite{Fischer2006infrared,LiPodzorov2007} Neither experiment shows any evidence for a second peak in the optical conductivity at 2000 cm$^{-1}$, which is therefore likely to be a deficiency of the SSH model we have used in our calculations. The measurement by Li {\em et al.}\cite{LiPodzorov2007} also does not show any evidence of a low-frequency rise in $\sigma(\omega)$, but the measurement by Fischer {\em et al.}\cite{Fischer2006infrared} does. The experimental evidence for a low frequency rise in $\sigma(\omega)$ caused by a long-time tail in $C(t)$ is therefore ambiguous.

\begin{figure}
    \centering
    \includegraphics{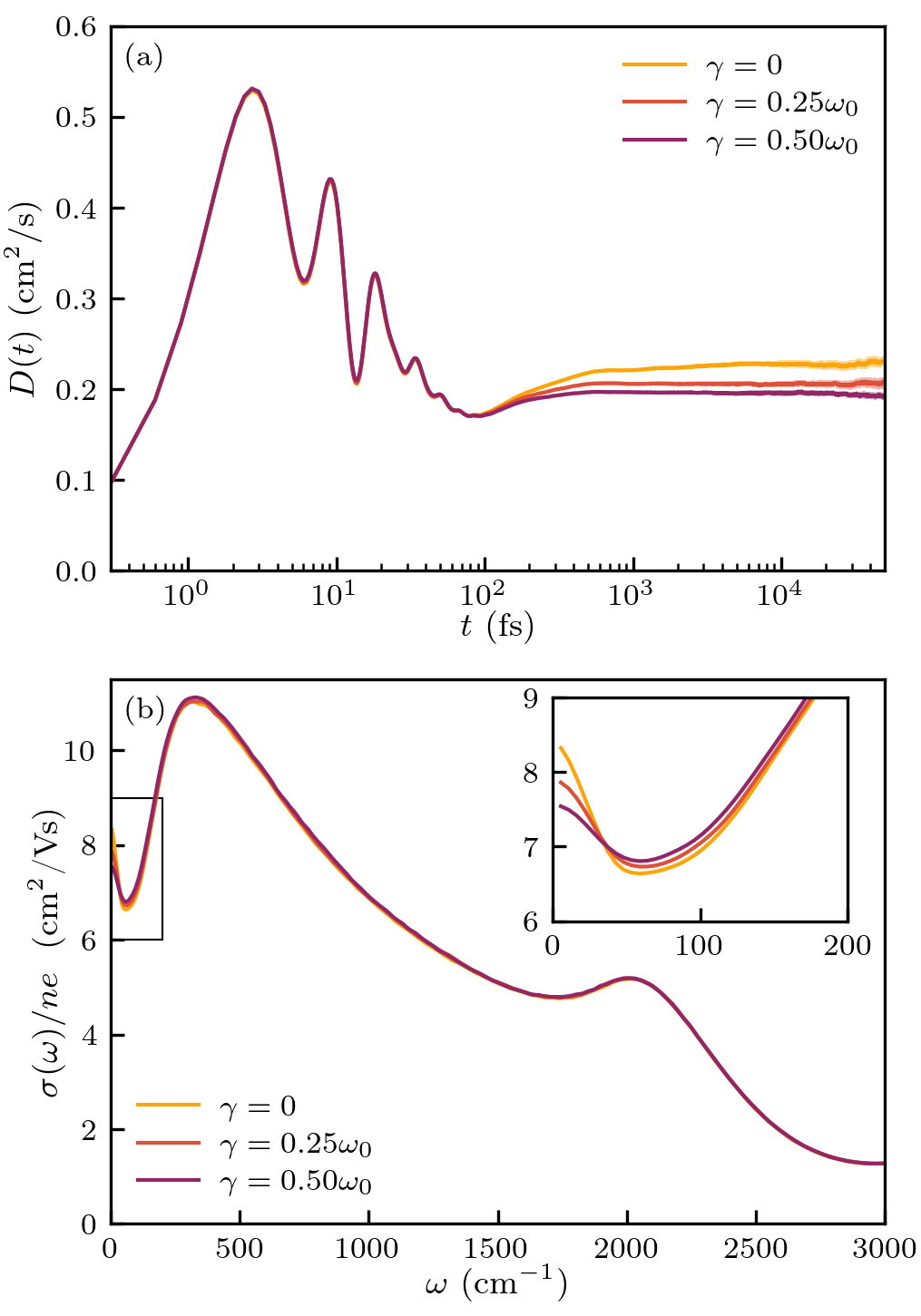}
    \caption{Impact of friction on the MASH dynamics. Increased friction leads to a reduction in (a) the long-time limit of the time-dependent diffusivity, and (b) the low-frequency limit of the optical conductivity.}
    \label{fig:friction}
\end{figure}

The low-frequency rise in the MASH calculation is only seen below $\omega_0$ ($\SI{40}{cm^{-1}}$), indicating that it is related to phonon motion. It is conceivable that the present model, which only includes a single-frequency phonon mode, is insufficient to represent the dynamics in a real material. A more realistic model would include a continuous density of phonon frequencies. For example, anharmonicity will effectively broaden the lineshape of the phonon spectrum, 
\begin{equation}
L(\omega)=\int_{-\infty}^{\infty} \rd t \,e^{-i\omega t} \langle q_j(0)q_j(t)\rangle. 
\end{equation}
A practical way to mimic this effect is to add Langevin friction to the dynamics, so that Eq.~\eqref{pdot} becomes \cite{nitzan2006book,Troisi2009disorder}
\begin{equation}
\dot{p}_j = F_j(q) - \gamma p_j + R_j(t)
\end{equation}
where $\gamma$ is a friction parameter and $R_j(t)$ is a random force that obeys $\langle R_j(t)R_j(0)\rangle = 2m\gamma\kBT \delta(t)$. In the underdamped regime ($\gamma < \omega_0$), $\gamma$ is approximately equal to the full width at half maximum of the lineshape $L(\omega)$. Based on the phonon spectrum in Fig.~1 of Ref.~\onlinecite{Troisi2007ehrenfest}, which is dominated by a single broad peak, we estimate a realistic friction to be on the order of $\gamma\approx \omega_0/4$. To explore the effect of friction on the MASH dynamics, Fig.~\ref{fig:friction} shows the results with $\gamma=0.25\,\omega_0$ and $\gamma=0.5\,\omega_0$. In comparison to the frictionless model, the only difference is in the dampening of the long-time dynamics and the associated low-frequency feature in the optical conductivity. Since other sources of dissipation may also damp out this feature, and since the amount of dissipation seen in an experiment is likely to depend on various factors including the quality of the crystalline sample, this may help to explain why the low-frequency feature is observed in one experiment in Fig.~\ref{fig:sigmaw}(b) and not in the other.

\begin{figure}
    \centering
    \includegraphics{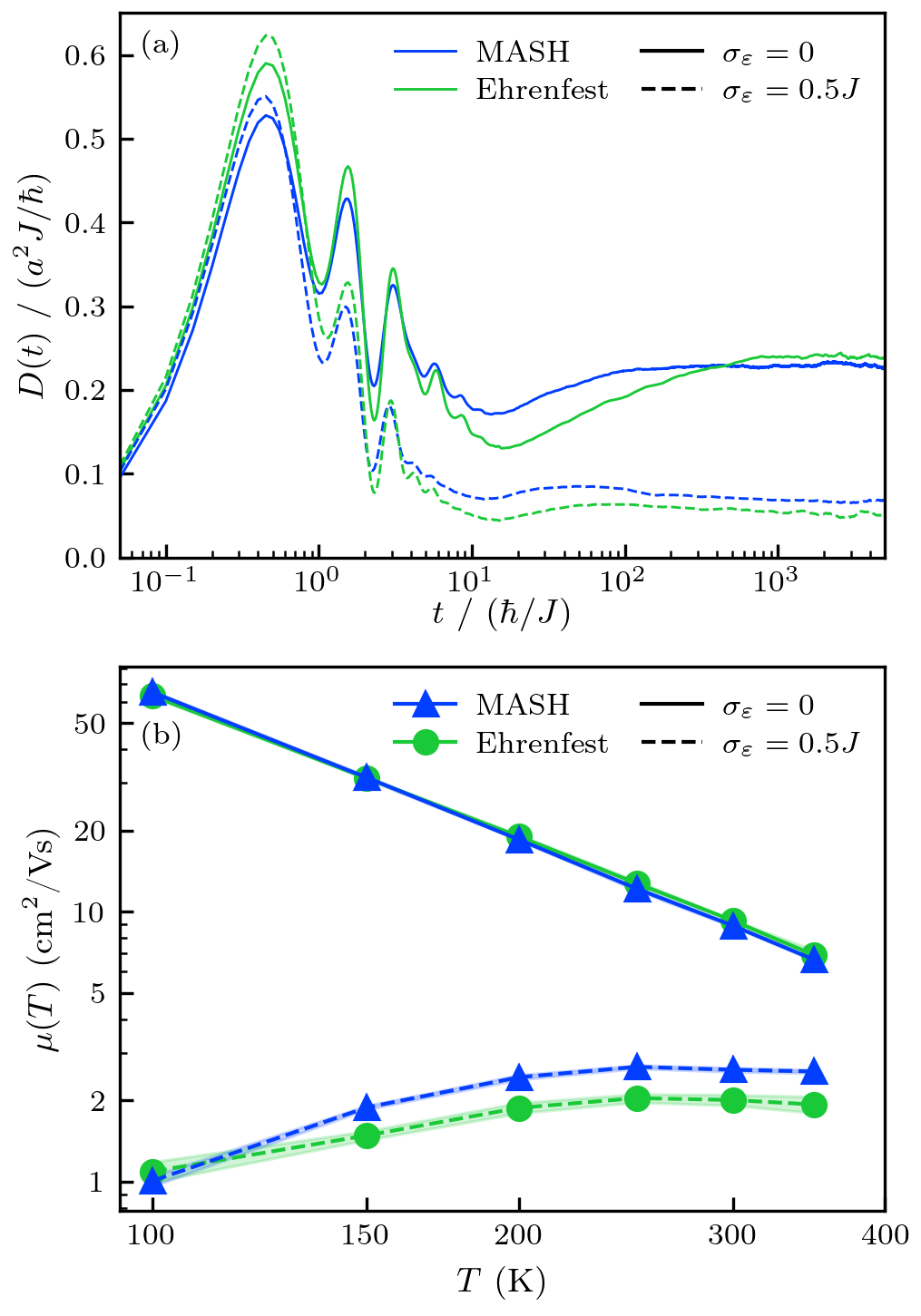}
    \caption{\Rev{Impact of static on-site disorder on the MASH and Ehrenfest dynamics. (a) Time-dependent diffusivities at $T=\SI{300}{K}$. (b) Temperature-dependent mobilities. Upon introducing static disorder, both methods predict a turnover in the mobility curve, but the precise shape of the turnover is different in the two calculations. (Here the shaded areas are one standard deviation error bars from the plateau regions of our diffusivity calculations.)}} 
    \label{fig:disorder}
\end{figure}

\rev{\subsection{Static disorder}
In real molecular crystals, further sources of disorder include lattice defects, impurities, and density fluctuations. A simple way to model such static disorder is to add a term $H_\mathrm{site} = \sum_j \varepsilon_j |j\rangle\langle j|$ to the Hamiltonian, where $\varepsilon_j$ is a time-independent on-site energy. For each trajectory, one samples each site energy independently from a Gaussian distribution with mean zero and standard deviation $\sigma_\varepsilon$, so that an average over many trajectories  averages over the disorder.}

\Rev{Figure~\ref{fig:disorder} shows the resulting time-dependent diffusivities at \SI{300}{K} and the temperature-dependent mobilities obtained using MASH and Ehrenfest dynamics for the case where $\sigma_{\varepsilon}=0.5\,J$. The results are compared with those for $\sigma_{\varepsilon}=0$ in Fig.~\ref{fig:Dt}. In a system where the site-to-site differences in $\varepsilon_j$ are larger than $\kBT$ ($0.235\,J$ at \SI{300}{K}) one might expect Ehrenfest dynamics to be less reliable, since it will lead to overpopulation of the high-energy sites. It is therefore interesting to note that we {\em do} see a difference between the Ehrenfest and MASH mobilities in Fig.~\ref{fig:disorder} when $\sigma_{\varepsilon}=0.5\,J$. Both methods predict a turnover in the temperature-dependence of the mobility, but the shapes of the two turnover curves are different, and we would expect this difference to become more pronounced with increasing $\sigma_{\varepsilon}$. This would be difficult to establish for the present rubrene model because increasing $\sigma_{\varepsilon}$ decreases the mobility and increases its relative statistical error, making the calculation more expensive. However, it is possible that the differences between Ehrenfest dynamics and MASH will become more apparent for models of other materials.}

\section{Conclusions} \label{sec:conclusions}
In this paper, we have presented a mixed quantum--classical method that can be used to simulate charge diffusion in molecular semiconductors at thermal equilibrium. The method solves the overheating problem of Ehrenfest dynamics and the CPA and it gives a well-defined long-time diffusivity without having to invoke a relaxation time approximation. \rev{In fact, it is the only mixed quantum-classical method we are aware of that is capable of simulating the thermal equilibrium dynamics of a quantum charge coupled to classical phonons in the diffusive regime -- the regime where the MASH diffusivity curve in Fig.~\ref{fig:Dt} has reached a plateau.} Since it rigorously preserves the mixed quantum--classical equilibrium distribution, one can also time average the correlation function along MASH trajectories to reduce the statistical error in the simulation. 

\rev{In spite of these formal arguments, it is interesting to note that the asymptotic mobilities obtained with Ehrenfest dynamics in Figs.~\ref{fig:Dt}, \ref{fig:muT}, and \ref{fig:disorder} are remarkably similar to the MASH results. One possible explanation could be that the feedback of the charge on the nuclear motion is comparatively weak in this system, so the choice of dynamical method is unimportant. This explanation is consistent with the relatively close agreement between the Ehrenfest and CPA (no backaction) results in Fig.~1, and with previous work on similar models of organic materials.\cite{Wang2011charge} Another possibility is that both methods are unphysical, but for different reasons. The formulation of MASH used here is not size-extensive and allows hops between uncoupled states, which may lead to an overestimated mobility. An alternative formulation of multi-state MASH has recently been proposed to address these issues\cite{Lawrence2024sizeconsistent} but that method is unlikely to be practical for a system with 100 or more sites. It is conceivable that the difference between Ehrenfest dynamics and the MASH method used here would be larger for systems with stronger electron-phonon coupling, and it will be interesting to explore this in future work.}

Like other methods that treat the phonons with classical variables, the present approach is only justified for low-frequency modes with a high thermal excitation. This is why we have not considered on-site Holstein modes, since their frequencies are typically high in energy compared to $\kBT$ and the classical treatment would neglect their zero-point energy. Such modes can significantly reduce the mobility.\cite{Dettmann2023killer,Knepp2024boltzmann} However, previous work suggests that the main effect can be included through the Lang--Firsov polaron transformation,\cite{Fetherolf2020prx,Nematiaram2020perspective} which effectively reduces the off-diagonal transfer integral (narrows the band width) and its fluctuations (the charge--phonon coupling). One can then use the present scheme on the renormalized model, with the low-frequency modes still treated classically.

In the present calculations, the MASH mobility is 30--60\,\% higher than that of the RTA with the default choice of its phenomenological relaxation time parameter ($\tau=\omega_0^{-1}$).\cite{Ciuchi2011} The difference is smaller when we include friction in the phonon motion as a simple model of a more realistic (anharmonic) system. 
Since the RTA is based on a static calculation, it is considerably cheaper than MASH, and we expect it to remain the most popular way to estimate the charge mobilities of different materials.

We emphasize that in MASH, the time-evolution of the complex vector $c$ is not intended to represent the motion of a physical wavefunction. It is merely a set of phase-space variables for the electronic degrees of freedom. If we nevertheless attempt to make a physical interpretation, we find it noteworthy that the method does not require any ``decoherence corrections'' or ``wavefunction collapse'' of $c$. This stands in contrast to the standard picture of a charge that undergoes transient localization between short spurts of coherent evolution on the length scale of a few molecules. The vector $c$ evolves coherently and remains delocalised throughout our equilibrium simulation, yet the classical degrees of freedom experience the force of a local charge because the relevant adiabatic states are local (at least in the parameter regime we have considered here). 

\rev{One final comment concerns the low-frequency rise in the optical conductivity we have found in our MASH calculations in Figs.~\ref{fig:sigmaw} and \ref{fig:friction}, which has also been seen in previous calculations using the CPA.\cite{Fetherolf2020prx,Fetherolf2023prb} This could be an artefact of these quantum-classical methods, an artefact of the SSH model, or something real. To help decide which, it would be useful to have an exact quantum mechanical benchmark calculation for the present model, which might be possible (for example) using the hierarchical equations of motion methodology developed in Ref.~\onlinecite{Dunn2019heom}. However, since the low frequency rise in our $\sigma(\omega)$ comes from a long-time tail in our $C(t)$, one would have to run the quantum dynamics for a sufficiently long time (at least out to $t=100\,\hbar/J$) to confirm or preclude the existence of this tail, which would be difficult to do for a system with $O(100)$ sites. (Numerical analytic continuation of the imaginary time veclocity autocorrelation function provides an alternative approach that has recently become popular in this field.\cite{DeFilippis2015qmc,Jankovic2022prb,Ostmeyer2023hmc} However, this is subject to the objection that the imaginary time correlation function does not contain any useful information about the real time dynamics beyond the thermal time $\beta\hbar$,\footnote{One way to see this is to note that the imaginary time correlation can be written as an integral over the real time correlation function with a kernel that decays as $e^{-2\pi t/\beta\hbar}$. See Eq.~(22) of Ref.~\onlinecite{Habershon2007meac}}\nocite{Habershon2007meac} which is a mere $4.3\,\hbar/J$ in the present calculations at \SI{300}{K}.)} It would be even more useful if  experimentalists were able to reach a consensus about the behaviour of the optical conductivity in the low frequency region of our Fig.~\ref{fig:sigmaw}, which is the heart of the matter. 

\rev{\section*{Supplementary Material}
The supplementary material discusses finite-size effects in Ehrenfest and CPA calculations and presents the convergence tests we have performed for these methods and MASH as a function of $N$.}

\section*{Acknowledgements}
The authors would like to thank William Barford, David Reichman, and Alessandro Troisi for insightful discussions. Johan Runeson was funded by a mobility fellowship from the Swiss National Science Foundation and supported by a junior research fellowship from Wadham College, Oxford.

\section*{Author declarations}

\subsection*{Conflict of interest}

The authors have no conflicts to disclose.

\section*{Data availability}

The data that support the findings of this study are available within the article.
%\revb{A source code containing an implementation of the present algorithm and relevant examples is publicly available at the Github repository \texttt{github.com/jruneson/multimash}.}

\appendix
\section{Derivation of the MASH Methodology}

The main new result in this paper is the expression in Eq.~\eqref{CMASH} for the velocity autocorrelation function of the charge in the quantum-classical limit where the phonons are treated classically. Here we provide a derivation of this expression in three stages: Section~1 summarises the overall argument, Section~2 derives an identity that it hinges on, and Section~3 works out an integral that appears in the result. 
\\

\subsection{Overall Argument}

We begin by writing the quantum-classical limit of Eq.~\eqref{Ct} as
\begin{align}
C(t) &= {1\over 2}\int {{\rm d}p\,{\rm d}q}\,
{\rm tr}_{\rm el}\Bigl[\hat{\rho}(p,q)\nonumber\\ &\times \Bigl(\hat{v}(q)\hat{v}(p,q,t)+\hat{v}(p,q,t)\hat{v}(q)\Bigr)\Bigr],
\label{A1}
\end{align}
where
\begin{equation}
\hat{\rho}(p,q) = {e^{-\beta\hat{H}(p,q)}\over \int {\rm d}p\,{\rm d}q \,{\rm tr}_{\rm el}\Bigl[e^{-\beta \hat{H}(p,q)}\Bigr]}\label{A2}
\end{equation}
is the thermal quantum-classical density operator, $\hat{v}(q)$ is the velocity operator defined in Eq.~\eqref{vop}, and $\hat{v}(p,q,t)$ is the same operator evolved to time $t$.

\def\br{\bar{\rho}}
\def\bv{\bar{v}}

Now $\hat{\rho}(p,q)$ is diagonal in the adiabatic basis at $q$, and the diagonal matrix elements of $\hat{v}(q)$ are zero in this basis. For operators of this form, Section~2 shows that the trace of the operator product in Eq.~\eqref{A1} can be written as
\begin{align}
{\rm tr}_{\rm el}\Bigl[&\hat{\rho}(p,q)\Bigl(\hat{v}(q)\hat{v}(p,q,t)+\hat{v}(p,q,t)\hat{v}(q)\Bigr)\Bigr]\nonumber\\
&= {1\over\gamma}\int_{|c|=1} {{\rm d}c}\, \br(p,q,c)\bv(q,c)\bv(p,q,c,t), \label{A3}
\end{align}
where
\begin{equation}
\gamma = \int_{|c|=1} {{\rm d}c}\,\, \Theta_a(c)|c_a|^2|c_b|^2 \label{gamma}
\end{equation}
with $b\not=a$ and
\begin{equation}
\br(p,q,c) = \sum_{a=1}^N \langle a(q)|\hat{\rho}(p,q)|a(q)\rangle \Theta_a(c)\label{brpqc}
\end{equation}
\begin{equation}
\bv(q,c) = \sum_{a=1}^N\sum_{b\not=a}^N \langle a(q)|\hat{v}(q)|b(q)\rangle \Theta_{ab}(c)c_a^*c_b\label{bvqc}
\end{equation}
\begin{equation}
\bv(p,q,c,t) = \sum_{a=1}^N\sum_{b=1}^N \langle a(q)|\hat{v}(p,q,t)|b(q)\rangle \Theta_{ab}(c) c_a^*c_b \label{bvpqct}
\end{equation}
with $\Theta_{ab}(c) = \Theta_a(c)+\Theta_b(c)$. 

The next stage of the argument is to replace $\bv(p,q,c,t)$ with $\bv(q_t,c_t)$, where $(p_t,q_t,c_t)$ is evolved from $(p,q,c)$ using the dynamics in Eqs.~\eqref{cdot}--\eqref{pdot} with the force in Eq.~\eqref{Fmash}. In the case of static disorder, where $q_t=q$ and the adiabatic populations $P_a=|c_a|^2$ are constants of the motion so $\Theta_{ab}(c)$ is independent of time, this replacement is exact, because Eq.~\eqref{cdot} is the time-dependent Schr\"odinger equation for the evolution of $c$ and writing $\hat{v}(p,q,t) = e^{+i\hat{V}(q)t/\hbar}\hat{v}(q)e^{-i\hat{V}(q)t/\hbar}$ in Eq.~\eqref{bvpqct} is equivalent to replacing $c$ with $c_t = e^{-i\hat{V}(q)t/\hbar}c$ in Eq.~\eqref{bvqc}. In the more general case where $q_t$ depends on time, the replacement $\bv(p,q,c,t)\to \bv(q_t,c_t)$ is no longer a quantum mechanical identity, but it is still consistent with the quantum-classical dynamics that is used to calculate $q_t$ and $c_t$ from $p$, $q$, $c$, and $t$.

The last stage of the argument is to note that $\br(p,q,c)$ in Eq.~\eqref{brpqc} can be written in terms of $\rho(p,q,c)$ in Eq.~\eqref{rhopqc} as
\begin{equation}
\br(p,q,c) = {\cal N}\rho(p,q,c)
\end{equation}
where\cite{Runeson2023mash}
\begin{equation}
{\cal N} = \int_{|c|=1} {\rm d}c\,\,\Theta_a(c) = \frac{2\pi^N}{N!}.
\end{equation}
Combining this with the factor of $1/2$ in Eq.~\eqref{A1} and the factor of $1/\gamma$ in Eq.~\eqref{A3},
replacing $\br(p,q,c)$ with ${\cal N}\rho(p,q,c)$ and $\bv(p,q,c,t)$ with $\bv(q_t,c_t)$, and rewriting $(p,q,c)$ as $(p_0,q_0,c_0)$ to emphasise what is evolved and what is not, we obtain
\begin{equation}
C(t) = {1\over 2\Gamma}\langle\bar{v}(q_0,c_0)\bar{v}(q_t,c_t)\rangle, \label{Cbar}
\end{equation}
where the average $\langle\cdots\rangle$ is as defined in Eq.~\eqref{MASHaverage} and
\begin{equation}
\Gamma = \int_{|c|=1} \frac{{\rm d}c}{\cal N}\,\,\Theta_a(c) |c_a|^2 |c_b|^2 \label{Gamma}
\end{equation}
%\Gamma = \frac{\int_{|c|=1} {\rm %d}c\,\,\Theta_a(c)|c_a|^2|c_b|^2}{\int_{|c|=1} {\rm d}c%\,\,\Theta_a(c)}\label{Gamma}
%\end{equation}
with $b\not=a$. This constant is worked out in Section~3. Finally, defining $v(q,c)=\sqrt{1/2\Gamma}\,\bv(q,c)$ as in Eq.~\eqref{vqc}, we see that Eq.~\eqref{Cbar} becomes $C(t) = \langle v(q_0,c_0)v(q_t,c_t)\rangle$, which is Eq.~\eqref{CMASH}.

\subsection{A Trace Identity}

Let $A$, $B$, and $C$ be the matrix representations of the operators $\hat{\rho}(p,q)$, $\hat{v}(q)$, and $\hat{v}(p,q,t)$ in the locally adiabatic basis at $q$. Since $A$ only has non-zero diagonal elements and $B$ only has non-zero off-diagonal elements in this basis,  the left-hand side of Eq.~\eqref{A3} is
\begin{equation}
{\rm tr}\Bigl[A\Bigl(BC+CB\Bigr)\Bigr] = \sum_{a=1}^N\sum_{b\not=a}^N A_{aa}\Bigl(B_{ab}C_{ba}+C_{ab}B_{ba}\Bigr).
\end{equation}
Our goal is to show that this is proportional to
\begin{equation}
\int_{|c|=1} {\rm d}c\, A(c)B(c)C(c)\label{ansatz}
\end{equation}
where
\begin{equation}
A(c) = \sum_{a=1}^N A_{aa}\Theta_a(c)\label{Ac}
\end{equation}
\begin{equation}
B(c) = \sum_{b=1}^N\sum_{b'\not=b}^N B_{bb'}\Theta_{bb'}(c) c_b^*c_{b'}\label{Bc}
\end{equation}
\begin{equation}
C(c) = \sum_{c=1}^N\sum_{c'=1}^N C_{cc'}\Theta_{cc'}(c) c_c^*c_{c'}\label{Cc}.
\end{equation}

Substituting Eqs.~\eqref{Ac}--\eqref{Cc} into Eq.~\eqref{ansatz} gives the rather lengthy expression
\begin{align}
\int_{|c|=1} &{\rm d}c\, A(c)B(c)C(c)=
\sum_{a=1}^N\sum_{b=1}^N\sum_{b'\not=b}^N\sum_{c=1}^N\sum_{c'=1}^NA_{aa}B_{bb'}C_{cc'}\nonumber\\
%\sum_a\sum_{bb'}\sprime\sum_{cc'} A_{aa}B_{bb'}C_{cc'}\nonumber\\
&\times  \int_{|c|=1} {\rm d}c\,\Theta_a(c)\Theta_{bb'}(c)\Theta_{cc'}(c)c_b^*c_{b'}c_c^*c_{c'}.
\end{align}
However, the integral on the right-hand side of this expression will only be non-zero if the phases cancel in the product $c_b^*c_{b'}c_c^*c_{c'}$. Since $b'\not=b$, phase cancellation can only happen if $b'=c$ and $c'=b$, which gives
\begin{align}
\int_{|c|=1} &{\rm d}c\, A(c)B(c)C(c)=
\sum_{a=1}^N\sum_{b=1}^N\sum_{c\not=b}^N A_{aa}B_{bc}C_{cb}\nonumber\\
%\sum_a\sum_{bc}\sprime A_{aa}B_{bc}C_{cb}\nonumber\\
&\times  \int_{|c|=1} {\rm d}c\,\Theta_a(c)\Theta_{bc}(c)|c_b|^2|c_c|^2.
\end{align}
The product of the two remaining projection operators is $\Theta_a(c)\Theta_{bc}(c)=\Theta_a(c)[\Theta_b(c)+\Theta_c(c)] = \Theta_a(c)[\delta_{ab}+\delta_{ac}]$, so this further simplifies to
\begin{equation}
\int_{|c|=1} {\rm d}c\, A(c)B(c)C(c)= \gamma\sum_{a=1}^N\sum_{b\not=a}^N A_{aa}\Bigl(B_{ab}C_{ba}+C_{ab}B_{ba}\Bigr)
%\int_{|c|=1} {\rm d}c\, A(c)B(c)C(c)= \gamma\sum_{ab}\sprime A_{aa}\Bigl(B_{ab}C_{ba}+C_{ab}B_{ba}\Bigr)
\end{equation}
where $\gamma$ is as defined in Eq.~\eqref{gamma}. Hence
\begin{equation}
{\rm tr}\Bigl[A\Bigl(BC+CB\Bigr)\Bigr] ={1\over\gamma}\int_{|c|=1} {\rm d}c\, A(c)B(c)C(c),\label{traceformula}
\end{equation}
which is the result we have used in Eq.~\eqref{A3}.

\subsection{A Surface Integral}

The final task is to evaluate $\Gamma$ in Eq.~\eqref{Gamma}. This can be interpreted as the expectation value $E[P_aP_b]$ of the product of two coordinates $P_a=|c_a|^2$ and $P_b=|c_b|^2$ on an $N$-simplex, where $P_a$ is the largest coordinate and $P_b$ is any one of the remaining $N-1$, all of which will give the same result for $E[P_aP_b]$ by symmetry:
\begin{equation}
\Gamma = E[P_aP_b] \equiv {1\over {N-1}}\Bigl(E[P_a]-E[P_a^2]\Bigr).
\end{equation} 

The first problem is thus to calculate $E[P_a]$ -- the expectation value of the largest coordinate on an $N$-simplex. As we have recently explained elsewhere,\cite{Runeson2023mash} this can be done by writing $P_a=\sum_{n=1}^N\tilde{P}_n/n$, where the $\tilde{P}_n$'s are the coordinates of an auxiliary simplex. Since $E[\tilde{P}_n]=1/N$, this immediately gives
\begin{equation}
E[P_a] = {H_N\over N},
\end{equation}
where $H_N=\sum_{n} 1/n$. It also gives
\begin{align}
E[P_a^2] &= \sum_{n=1}^N\sum_{n'=1}^N {E[\tilde{P}_n\tilde{P}_{n'}]\over nn'}\nonumber\\
               &= {1\over N(N+1)}\left[\sum_{n=1}^N\sum_{n'=1}^N {1\over nn'}+\sum_{n=1}^N {1\over n^2}\right]\nonumber\\
               &= {H_N^2+G_N\over N(N+1)},
\end{align}
where $G_N = \sum_{n=1}^N 1/n^2$ and we have used the formula
\begin{equation}
E[\tilde{P}_n\tilde{P}_{n'}] = {1+\delta_{nn'}\over N(N+1)}
\end{equation}
for the expectation value of a product of two coordinates on a simplex. Hence
\begin{equation}
\Gamma = {H_N\over N(N-1)}-{H_N^2+G_N\over (N+1)N(N-1)},
\end{equation}
which is Eq.~\eqref{gammaN} of the text.

\bibliography{runerefs.bib}

%merlin.mbs aipnum4-1.bst 2010-07-25 4.21a (PWD, AO, DPC) hacked
%Control: key (0)
%Control: author (8) initials jnrlst
%Control: editor formatted (1) identically to author
%Control: production of article title (0) allowed
%Control: page (1) range
%Control: year (1) truncated
%Control: production of eprint (-1) disabled
\begin{thebibliography}{62}%
\makeatletter
\providecommand \@ifxundefined [1]{%
 \@ifx{#1\undefined}
}%
\providecommand \@ifnum [1]{%
 \ifnum #1\expandafter \@firstoftwo
 \else \expandafter \@secondoftwo
 \fi
}%
\providecommand \@ifx [1]{%
 \ifx #1\expandafter \@firstoftwo
 \else \expandafter \@secondoftwo
 \fi
}%
\providecommand \natexlab [1]{#1}%
\providecommand \enquote  [1]{``#1''}%
\providecommand \bibnamefont  [1]{#1}%
\providecommand \bibfnamefont [1]{#1}%
\providecommand \citenamefont [1]{#1}%
\providecommand \href@noop [0]{\@secondoftwo}%
\providecommand \href [0]{\begingroup \@sanitize@url \@href}%
\providecommand \@href[1]{\@@startlink{#1}\@@href}%
\providecommand \@@href[1]{\endgroup#1\@@endlink}%
\providecommand \@sanitize@url [0]{\catcode `\\12\catcode `\$12\catcode
  `\&12\catcode `\#12\catcode `\^12\catcode `\_12\catcode `\%12\relax}%
\providecommand \@@startlink[1]{}%
\providecommand \@@endlink[0]{}%
\providecommand \url  [0]{\begingroup\@sanitize@url \@url }%
\providecommand \@url [1]{\endgroup\@href {#1}{\urlprefix }}%
\providecommand \urlprefix  [0]{URL }%
\providecommand \Eprint [0]{\href }%
\providecommand \doibase [0]{http://dx.doi.org/}%
\providecommand \selectlanguage [0]{\@gobble}%
\providecommand \bibinfo  [0]{\@secondoftwo}%
\providecommand \bibfield  [0]{\@secondoftwo}%
\providecommand \translation [1]{[#1]}%
\providecommand \BibitemOpen [0]{}%
\providecommand \bibitemStop [0]{}%
\providecommand \bibitemNoStop [0]{.\EOS\space}%
\providecommand \EOS [0]{\spacefactor3000\relax}%
\providecommand \BibitemShut  [1]{\csname bibitem#1\endcsname}%
\let\auto@bib@innerbib\@empty
%</preamble>
\bibitem [{\citenamefont {Fratini}\ \emph {et~al.}(2020)\citenamefont
  {Fratini}, \citenamefont {Nikolka}, \citenamefont {Salleo}, \citenamefont
  {Schweicher},\ and\ \citenamefont {Sirringhaus}}]{Fratini2020}%
  \BibitemOpen
  \bibfield  {author} {\bibinfo {author} {\bibfnamefont {S.}~\bibnamefont
  {Fratini}}, \bibinfo {author} {\bibfnamefont {M.}~\bibnamefont {Nikolka}},
  \bibinfo {author} {\bibfnamefont {A.}~\bibnamefont {Salleo}}, \bibinfo
  {author} {\bibfnamefont {G.}~\bibnamefont {Schweicher}}, \ and\ \bibinfo
  {author} {\bibfnamefont {H.}~\bibnamefont {Sirringhaus}},\ }\bibfield
  {title} {\enquote {\bibinfo {title} {Charge transport in high-mobility
  conjugated polymers and molecular semiconductors},}\ }\href {\doibase
  10.1038/s41563-020-0647-2} {\bibfield  {journal} {\bibinfo  {journal} {Nat.
  Mater.}\ }\textbf {\bibinfo {volume} {19}},\ \bibinfo {pages} {491--502}
  (\bibinfo {year} {2020})}\BibitemShut {NoStop}%
\bibitem [{\citenamefont {Ghosh}\ and\ \citenamefont
  {Spano}(2020)}]{Ghosh2020account}%
  \BibitemOpen
  \bibfield  {author} {\bibinfo {author} {\bibfnamefont {R.}~\bibnamefont
  {Ghosh}}\ and\ \bibinfo {author} {\bibfnamefont {F.~C.}\ \bibnamefont
  {Spano}},\ }\bibfield  {title} {\enquote {\bibinfo {title} {Excitons and
  polarons in organic materials},}\ }\href {\doibase
  10.1021/acs.accounts.0c00349} {\bibfield  {journal} {\bibinfo  {journal}
  {Acc. Chem. Res.}\ }\textbf {\bibinfo {volume} {53}},\ \bibinfo {pages}
  {2201--2211} (\bibinfo {year} {2020})}\BibitemShut {NoStop}%
\bibitem [{\citenamefont {Oberhofer}, \citenamefont {Reuter},\ and\
  \citenamefont {Blumberger}(2017)}]{Oberhofer2017}%
  \BibitemOpen
  \bibfield  {author} {\bibinfo {author} {\bibfnamefont {H.}~\bibnamefont
  {Oberhofer}}, \bibinfo {author} {\bibfnamefont {K.}~\bibnamefont {Reuter}}, \
  and\ \bibinfo {author} {\bibfnamefont {J.}~\bibnamefont {Blumberger}},\
  }\bibfield  {title} {\enquote {\bibinfo {title} {Charge transport in
  molecular materials: An assessment of computational methods},}\ }\href
  {\doibase 10.1021/acs.chemrev.7b00086} {\bibfield  {journal} {\bibinfo
  {journal} {Chem. Rev.}\ }\textbf {\bibinfo {volume} {117}},\ \bibinfo {pages}
  {10319--10357} (\bibinfo {year} {2017})}\BibitemShut {NoStop}%
\bibitem [{\citenamefont {Fratini}, \citenamefont {Mayou},\ and\ \citenamefont
  {Ciuchi}(2016)}]{Fratini2016}%
  \BibitemOpen
  \bibfield  {author} {\bibinfo {author} {\bibfnamefont {S.}~\bibnamefont
  {Fratini}}, \bibinfo {author} {\bibfnamefont {D.}~\bibnamefont {Mayou}}, \
  and\ \bibinfo {author} {\bibfnamefont {S.}~\bibnamefont {Ciuchi}},\
  }\bibfield  {title} {\enquote {\bibinfo {title} {The transient localization
  scenario for charge transport in crystalline organic materials},}\ }\href
  {\doibase https://doi.org/10.1002/adfm.201502386} {\bibfield  {journal}
  {\bibinfo  {journal} {Adv. Funct. Mater.}\ }\textbf {\bibinfo {volume}
  {26}},\ \bibinfo {pages} {2292--2315} (\bibinfo {year} {2016})}\BibitemShut
  {NoStop}%
\bibitem [{\citenamefont {Nematiaram}\ and\ \citenamefont
  {Troisi}(2020)}]{Nematiaram2020perspective}%
  \BibitemOpen
  \bibfield  {author} {\bibinfo {author} {\bibfnamefont {T.}~\bibnamefont
  {Nematiaram}}\ and\ \bibinfo {author} {\bibfnamefont {A.}~\bibnamefont
  {Troisi}},\ }\bibfield  {title} {\enquote {\bibinfo {title} {{Modeling charge
  transport in high-mobility molecular semiconductors: Balancing electronic
  structure and quantum dynamics methods with the help of experiments}},}\
  }\href {\doibase 10.1063/5.0008357} {\bibfield  {journal} {\bibinfo
  {journal} {J. Chem. Phys.}\ }\textbf {\bibinfo {volume} {152}},\ \bibinfo
  {pages} {190902} (\bibinfo {year} {2020})}\BibitemShut {NoStop}%
\bibitem [{\citenamefont {Giannini}\ \emph {et~al.}(2023)\citenamefont
  {Giannini}, \citenamefont {Di~Virgilio}, \citenamefont {Bardini},
  \citenamefont {Hausch}, \citenamefont {Geuchies}, \citenamefont {Zheng},
  \citenamefont {Volpi}, \citenamefont {Elsner}, \citenamefont {Broch},
  \citenamefont {Geerts}, \citenamefont {Schreiber}, \citenamefont
  {Schweicher}, \citenamefont {Wang}, \citenamefont {Blumberger}, \citenamefont
  {Bonn},\ and\ \citenamefont {Beljonne}}]{Giannini2023natmat}%
  \BibitemOpen
  \bibfield  {author} {\bibinfo {author} {\bibfnamefont {S.}~\bibnamefont
  {Giannini}}, \bibinfo {author} {\bibfnamefont {L.}~\bibnamefont
  {Di~Virgilio}}, \bibinfo {author} {\bibfnamefont {M.}~\bibnamefont
  {Bardini}}, \bibinfo {author} {\bibfnamefont {J.}~\bibnamefont {Hausch}},
  \bibinfo {author} {\bibfnamefont {J.~J.}\ \bibnamefont {Geuchies}}, \bibinfo
  {author} {\bibfnamefont {W.}~\bibnamefont {Zheng}}, \bibinfo {author}
  {\bibfnamefont {M.}~\bibnamefont {Volpi}}, \bibinfo {author} {\bibfnamefont
  {J.}~\bibnamefont {Elsner}}, \bibinfo {author} {\bibfnamefont
  {K.}~\bibnamefont {Broch}}, \bibinfo {author} {\bibfnamefont {Y.~H.}\
  \bibnamefont {Geerts}}, \bibinfo {author} {\bibfnamefont {F.}~\bibnamefont
  {Schreiber}}, \bibinfo {author} {\bibfnamefont {G.}~\bibnamefont
  {Schweicher}}, \bibinfo {author} {\bibfnamefont {H.~I.}\ \bibnamefont
  {Wang}}, \bibinfo {author} {\bibfnamefont {J.}~\bibnamefont {Blumberger}},
  \bibinfo {author} {\bibfnamefont {M.}~\bibnamefont {Bonn}}, \ and\ \bibinfo
  {author} {\bibfnamefont {D.}~\bibnamefont {Beljonne}},\ }\bibfield  {title}
  {\enquote {\bibinfo {title} {Transiently delocalized states enhance hole
  mobility in organic molecular semiconductors},}\ }\href {\doibase
  10.1038/s41563-023-01664-4} {\bibfield  {journal} {\bibinfo  {journal} {Nat.
  Mater.}\ }\textbf {\bibinfo {volume} {22}},\ \bibinfo {pages} {1361--1369}
  (\bibinfo {year} {2023})}\BibitemShut {NoStop}%
\bibitem [{\citenamefont {Fratini}\ \emph {et~al.}(2017)\citenamefont
  {Fratini}, \citenamefont {Ciuchi}, \citenamefont {Mayou}, \citenamefont
  {de~Laissardi{\`e}re},\ and\ \citenamefont {Troisi}}]{Fratini2017natmat}%
  \BibitemOpen
  \bibfield  {author} {\bibinfo {author} {\bibfnamefont {S.}~\bibnamefont
  {Fratini}}, \bibinfo {author} {\bibfnamefont {S.}~\bibnamefont {Ciuchi}},
  \bibinfo {author} {\bibfnamefont {D.}~\bibnamefont {Mayou}}, \bibinfo
  {author} {\bibfnamefont {G.~T.}\ \bibnamefont {de~Laissardi{\`e}re}}, \ and\
  \bibinfo {author} {\bibfnamefont {A.}~\bibnamefont {Troisi}},\ }\bibfield
  {title} {\enquote {\bibinfo {title} {A map of high-mobility molecular
  semiconductors},}\ }\href {\doibase 10.1038/nmat4970} {\bibfield  {journal}
  {\bibinfo  {journal} {Nat. Mater.}\ }\textbf {\bibinfo {volume} {16}},\
  \bibinfo {pages} {998--1002} (\bibinfo {year} {2017})}\BibitemShut {NoStop}%
\bibitem [{\citenamefont {Nematiaram}\ \emph {et~al.}(2019)\citenamefont
  {Nematiaram}, \citenamefont {Ciuchi}, \citenamefont {Xie}, \citenamefont
  {Fratini},\ and\ \citenamefont {Troisi}}]{Nematiaram2019practical}%
  \BibitemOpen
  \bibfield  {author} {\bibinfo {author} {\bibfnamefont {T.}~\bibnamefont
  {Nematiaram}}, \bibinfo {author} {\bibfnamefont {S.}~\bibnamefont {Ciuchi}},
  \bibinfo {author} {\bibfnamefont {X.}~\bibnamefont {Xie}}, \bibinfo {author}
  {\bibfnamefont {S.}~\bibnamefont {Fratini}}, \ and\ \bibinfo {author}
  {\bibfnamefont {A.}~\bibnamefont {Troisi}},\ }\bibfield  {title} {\enquote
  {\bibinfo {title} {Practical computation of the charge mobility in molecular
  semiconductors using transient localization theory},}\ }\href {\doibase
  10.1021/acs.jpcc.8b11916} {\bibfield  {journal} {\bibinfo  {journal} {J.
  Phys. Chem. C}\ }\textbf {\bibinfo {volume} {123}},\ \bibinfo {pages}
  {6989--6997} (\bibinfo {year} {2019})}\BibitemShut {NoStop}%
\bibitem [{\citenamefont {Harrelson}\ \emph {et~al.}(2019)\citenamefont
  {Harrelson}, \citenamefont {Dantanarayana}, \citenamefont {Xie},
  \citenamefont {Koshnick}, \citenamefont {Nai}, \citenamefont {Fair},
  \citenamefont {Nuñez}, \citenamefont {Thomas}, \citenamefont {Murrey},
  \citenamefont {Hickner}, \citenamefont {Grey}, \citenamefont {Anthony},
  \citenamefont {Gomez}, \citenamefont {Troisi}, \citenamefont {Faller},\ and\
  \citenamefont {Moulé}}]{Harrelson2019probe}%
  \BibitemOpen
  \bibfield  {author} {\bibinfo {author} {\bibfnamefont {T.~F.}\ \bibnamefont
  {Harrelson}}, \bibinfo {author} {\bibfnamefont {V.}~\bibnamefont
  {Dantanarayana}}, \bibinfo {author} {\bibfnamefont {X.}~\bibnamefont {Xie}},
  \bibinfo {author} {\bibfnamefont {C.}~\bibnamefont {Koshnick}}, \bibinfo
  {author} {\bibfnamefont {D.}~\bibnamefont {Nai}}, \bibinfo {author}
  {\bibfnamefont {R.}~\bibnamefont {Fair}}, \bibinfo {author} {\bibfnamefont
  {S.~A.}\ \bibnamefont {Nuñez}}, \bibinfo {author} {\bibfnamefont {A.~K.}\
  \bibnamefont {Thomas}}, \bibinfo {author} {\bibfnamefont {T.~L.}\
  \bibnamefont {Murrey}}, \bibinfo {author} {\bibfnamefont {M.~A.}\
  \bibnamefont {Hickner}}, \bibinfo {author} {\bibfnamefont {J.~K.}\
  \bibnamefont {Grey}}, \bibinfo {author} {\bibfnamefont {J.~E.}\ \bibnamefont
  {Anthony}}, \bibinfo {author} {\bibfnamefont {E.~D.}\ \bibnamefont {Gomez}},
  \bibinfo {author} {\bibfnamefont {A.}~\bibnamefont {Troisi}}, \bibinfo
  {author} {\bibfnamefont {R.}~\bibnamefont {Faller}}, \ and\ \bibinfo {author}
  {\bibfnamefont {A.~J.}\ \bibnamefont {Moulé}},\ }\bibfield  {title}
  {\enquote {\bibinfo {title} {Direct probe of the nuclear modes limiting
  charge mobility in molecular semiconductors},}\ }\href {\doibase
  10.1039/C8MH01069B} {\bibfield  {journal} {\bibinfo  {journal} {Mater.
  Horiz.}\ }\textbf {\bibinfo {volume} {6}},\ \bibinfo {pages} {182--191}
  (\bibinfo {year} {2019})}\BibitemShut {NoStop}%
\bibitem [{\citenamefont {Landi}(2019)}]{Landi2019mobility}%
  \BibitemOpen
  \bibfield  {author} {\bibinfo {author} {\bibfnamefont {A.}~\bibnamefont
  {Landi}},\ }\bibfield  {title} {\enquote {\bibinfo {title} {Charge mobility
  prediction in organic semiconductors: Comparison of second-order cumulant
  approximation and transient localization theory},}\ }\href {\doibase
  10.1021/acs.jpcc.9b04252} {\bibfield  {journal} {\bibinfo  {journal} {J.
  Phys. Chem. C}\ }\textbf {\bibinfo {volume} {123}},\ \bibinfo {pages}
  {18804--18812} (\bibinfo {year} {2019})}\BibitemShut {NoStop}%
\bibitem [{\citenamefont {De~Filippis}\ \emph {et~al.}(2015)\citenamefont
  {De~Filippis}, \citenamefont {Cataudella}, \citenamefont {Mishchenko},
  \citenamefont {Nagaosa}, \citenamefont {Fierro},\ and\ \citenamefont
  {de~Candia}}]{DeFilippis2015qmc}%
  \BibitemOpen
  \bibfield  {author} {\bibinfo {author} {\bibfnamefont {G.}~\bibnamefont
  {De~Filippis}}, \bibinfo {author} {\bibfnamefont {V.}~\bibnamefont
  {Cataudella}}, \bibinfo {author} {\bibfnamefont {A.~S.}\ \bibnamefont
  {Mishchenko}}, \bibinfo {author} {\bibfnamefont {N.}~\bibnamefont {Nagaosa}},
  \bibinfo {author} {\bibfnamefont {A.}~\bibnamefont {Fierro}}, \ and\ \bibinfo
  {author} {\bibfnamefont {A.}~\bibnamefont {de~Candia}},\ }\bibfield  {title}
  {\enquote {\bibinfo {title} {Crossover from super- to subdiffusive motion and
  memory effects in crystalline organic semiconductors},}\ }\href {\doibase
  10.1103/PhysRevLett.114.086601} {\bibfield  {journal} {\bibinfo  {journal}
  {Phys. Rev. Lett.}\ }\textbf {\bibinfo {volume} {114}},\ \bibinfo {pages}
  {086601} (\bibinfo {year} {2015})}\BibitemShut {NoStop}%
\bibitem [{\citenamefont {Li}, \citenamefont {Ren},\ and\ \citenamefont
  {Shuai}(2021)}]{Shuai2021mps}%
  \BibitemOpen
  \bibfield  {author} {\bibinfo {author} {\bibfnamefont {W.}~\bibnamefont
  {Li}}, \bibinfo {author} {\bibfnamefont {J.}~\bibnamefont {Ren}}, \ and\
  \bibinfo {author} {\bibfnamefont {Z.}~\bibnamefont {Shuai}},\ }\bibfield
  {title} {\enquote {\bibinfo {title} {A general charge transport picture for
  organic semiconductors with nonlocal electron-phonon couplings},}\ }\href
  {\doibase 10.1038/s41467-021-24520-y} {\bibfield  {journal} {\bibinfo
  {journal} {Nat. Commun.}\ }\textbf {\bibinfo {volume} {12}},\ \bibinfo
  {pages} {4260} (\bibinfo {year} {2021})}\BibitemShut {NoStop}%
\bibitem [{\citenamefont {Li}, \citenamefont {Yan},\ and\ \citenamefont
  {Shi}(2024)}]{Shi2024finite}%
  \BibitemOpen
  \bibfield  {author} {\bibinfo {author} {\bibfnamefont {T.}~\bibnamefont
  {Li}}, \bibinfo {author} {\bibfnamefont {Y.}~\bibnamefont {Yan}}, \ and\
  \bibinfo {author} {\bibfnamefont {Q.}~\bibnamefont {Shi}},\ }\bibfield
  {title} {\enquote {\bibinfo {title} {{Is there a finite mobility for the one
  vibrational mode Holstein model? Implications from real time simulations}},}\
  }\href {\doibase 10.1063/5.0198107} {\bibfield  {journal} {\bibinfo
  {journal} {J. Chem. Phys.}\ }\textbf {\bibinfo {volume} {160}},\ \bibinfo
  {pages} {111102} (\bibinfo {year} {2024})}\BibitemShut {NoStop}%
\bibitem [{\citenamefont {Ostmeyer}\ \emph {et~al.}(2023)\citenamefont
  {Ostmeyer}, \citenamefont {Nematiaram}, \citenamefont {Troisi},\ and\
  \citenamefont {Buividovich}}]{Ostmeyer2023hmc}%
  \BibitemOpen
  \bibfield  {author} {\bibinfo {author} {\bibfnamefont {J.}~\bibnamefont
  {Ostmeyer}}, \bibinfo {author} {\bibfnamefont {T.}~\bibnamefont
  {Nematiaram}}, \bibinfo {author} {\bibfnamefont {A.}~\bibnamefont {Troisi}},
  \ and\ \bibinfo {author} {\bibfnamefont {P.}~\bibnamefont {Buividovich}},\
  }\href {https://arxiv.org/abs/2312.14914} {\enquote {\bibinfo {title}
  {First-principle quantum monte-carlo study of charge carrier mobility in
  organic molecular semiconductors},}\ } (\bibinfo {year} {2023}),\ \bibinfo
  {note} {arXiv:2312.14914 [cond-mat.mtrl-sci]}\BibitemShut {NoStop}%
\bibitem [{\citenamefont {Troisi}\ and\ \citenamefont
  {Orlandi}(2006)}]{Troisi2006}%
  \BibitemOpen
  \bibfield  {author} {\bibinfo {author} {\bibfnamefont {A.}~\bibnamefont
  {Troisi}}\ and\ \bibinfo {author} {\bibfnamefont {G.}~\bibnamefont
  {Orlandi}},\ }\bibfield  {title} {\enquote {\bibinfo {title}
  {Charge-transport regime of crystalline organic semiconductors: Diffusion
  limited by thermal off-diagonal electronic disorder},}\ }\href {\doibase
  10.1103/PhysRevLett.96.086601} {\bibfield  {journal} {\bibinfo  {journal}
  {Phys. Rev. Lett.}\ }\textbf {\bibinfo {volume} {96}},\ \bibinfo {pages}
  {086601} (\bibinfo {year} {2006})}\BibitemShut {NoStop}%
\bibitem [{\citenamefont {Parandekar}\ and\ \citenamefont
  {Tully}(2005)}]{Parandekar2005mixed}%
  \BibitemOpen
  \bibfield  {author} {\bibinfo {author} {\bibfnamefont {P.~V.}\ \bibnamefont
  {Parandekar}}\ and\ \bibinfo {author} {\bibfnamefont {J.~C.}\ \bibnamefont
  {Tully}},\ }\bibfield  {title} {\enquote {\bibinfo {title} {Mixed
  quantum-classical equilibrium},}\ }\href {\doibase 10.1063/1.1856460}
  {\bibfield  {journal} {\bibinfo  {journal} {J. Chem. Phys.}\ }\textbf
  {\bibinfo {volume} {122}},\ \bibinfo {pages} {094102} (\bibinfo {year}
  {2005})}\BibitemShut {NoStop}%
\bibitem [{\citenamefont {Ciuchi}, \citenamefont {Fratini},\ and\ \citenamefont
  {Mayou}(2011)}]{Ciuchi2011}%
  \BibitemOpen
  \bibfield  {author} {\bibinfo {author} {\bibfnamefont {S.}~\bibnamefont
  {Ciuchi}}, \bibinfo {author} {\bibfnamefont {S.}~\bibnamefont {Fratini}}, \
  and\ \bibinfo {author} {\bibfnamefont {D.}~\bibnamefont {Mayou}},\ }\bibfield
   {title} {\enquote {\bibinfo {title} {Transient localization in crystalline
  organic semiconductors},}\ }\href {\doibase 10.1103/PhysRevB.83.081202}
  {\bibfield  {journal} {\bibinfo  {journal} {Phys. Rev. B}\ }\textbf {\bibinfo
  {volume} {83}},\ \bibinfo {pages} {081202} (\bibinfo {year}
  {2011})}\BibitemShut {NoStop}%
\bibitem [{\citenamefont {Tully}(1990)}]{Tully1990hopping}%
  \BibitemOpen
  \bibfield  {author} {\bibinfo {author} {\bibfnamefont {J.~C.}\ \bibnamefont
  {Tully}},\ }\bibfield  {title} {\enquote {\bibinfo {title} {Molecular
  dynamics with electronic transitions},}\ }\href {\doibase 10.1063/1.459170}
  {\bibfield  {journal} {\bibinfo  {journal} {J.~Chem. Phys.}\ }\textbf
  {\bibinfo {volume} {93}},\ \bibinfo {pages} {1061--1071} (\bibinfo {year}
  {1990})}\BibitemShut {NoStop}%
\bibitem [{\citenamefont {Wang}\ and\ \citenamefont
  {Beljonne}(2013)}]{Wang2013flexible}%
  \BibitemOpen
  \bibfield  {author} {\bibinfo {author} {\bibfnamefont {L.}~\bibnamefont
  {Wang}}\ and\ \bibinfo {author} {\bibfnamefont {D.}~\bibnamefont
  {Beljonne}},\ }\bibfield  {title} {\enquote {\bibinfo {title} {Flexible
  surface hopping approach to model the crossover from hopping to band-like
  transport in organic crystals},}\ }\href {\doibase 10.1021/jz400871j}
  {\bibfield  {journal} {\bibinfo  {journal} {J. Phys. Chem. Lett.}\ }\textbf
  {\bibinfo {volume} {4}},\ \bibinfo {pages} {1888--1894} (\bibinfo {year}
  {2013})}\BibitemShut {NoStop}%
\bibitem [{\citenamefont {Wang}, \citenamefont {Prezhdo},\ and\ \citenamefont
  {Beljonne}(2015)}]{Wang2015transport}%
  \BibitemOpen
  \bibfield  {author} {\bibinfo {author} {\bibfnamefont {L.}~\bibnamefont
  {Wang}}, \bibinfo {author} {\bibfnamefont {O.~V.}\ \bibnamefont {Prezhdo}}, \
  and\ \bibinfo {author} {\bibfnamefont {D.}~\bibnamefont {Beljonne}},\
  }\bibfield  {title} {\enquote {\bibinfo {title} {Mixed quantum-classical
  dynamics for charge transport in organics},}\ }\href {\doibase
  10.1039/C5CP00485C} {\bibfield  {journal} {\bibinfo  {journal} {Phys. Chem.
  Chem. Phys.}\ }\textbf {\bibinfo {volume} {17}},\ \bibinfo {pages}
  {12395--12406} (\bibinfo {year} {2015})}\BibitemShut {NoStop}%
\bibitem [{\citenamefont {Giannini}\ \emph {et~al.}(2019)\citenamefont
  {Giannini}, \citenamefont {Carof}, \citenamefont {Ellis}, \citenamefont
  {Yang}, \citenamefont {Ziogos}, \citenamefont {Ghosh},\ and\ \citenamefont
  {Blumberger}}]{Giannini2019}%
  \BibitemOpen
  \bibfield  {author} {\bibinfo {author} {\bibfnamefont {S.}~\bibnamefont
  {Giannini}}, \bibinfo {author} {\bibfnamefont {A.}~\bibnamefont {Carof}},
  \bibinfo {author} {\bibfnamefont {M.}~\bibnamefont {Ellis}}, \bibinfo
  {author} {\bibfnamefont {H.}~\bibnamefont {Yang}}, \bibinfo {author}
  {\bibfnamefont {O.~G.}\ \bibnamefont {Ziogos}}, \bibinfo {author}
  {\bibfnamefont {S.}~\bibnamefont {Ghosh}}, \ and\ \bibinfo {author}
  {\bibfnamefont {J.}~\bibnamefont {Blumberger}},\ }\bibfield  {title}
  {\enquote {\bibinfo {title} {Quantum localization and delocalization of
  charge carriers in organic semiconducting crystals},}\ }\href {\doibase
  10.1038/s41467-019-11775-9} {\bibfield  {journal} {\bibinfo  {journal} {Nat.
  Commun.}\ }\textbf {\bibinfo {volume} {10}},\ \bibinfo {pages} {3843}
  (\bibinfo {year} {2019})}\BibitemShut {NoStop}%
\bibitem [{\citenamefont {Xie}\ \emph {et~al.}(2020)\citenamefont {Xie},
  \citenamefont {Holub}, \citenamefont {Kubař},\ and\ \citenamefont
  {Elstner}}]{Xie2020semiconductor}%
  \BibitemOpen
  \bibfield  {author} {\bibinfo {author} {\bibfnamefont {W.}~\bibnamefont
  {Xie}}, \bibinfo {author} {\bibfnamefont {D.}~\bibnamefont {Holub}}, \bibinfo
  {author} {\bibfnamefont {T.}~\bibnamefont {Kubař}}, \ and\ \bibinfo {author}
  {\bibfnamefont {M.}~\bibnamefont {Elstner}},\ }\bibfield  {title} {\enquote
  {\bibinfo {title} {Performance of mixed quantum-classical approaches on
  modeling the crossover from hopping to bandlike charge transport in organic
  semiconductors},}\ }\href {\doibase 10.1021/acs.jctc.9b01271} {\bibfield
  {journal} {\bibinfo  {journal} {J. Chem. Theory Comput.}\ }\textbf {\bibinfo
  {volume} {16}},\ \bibinfo {pages} {2071--2084} (\bibinfo {year}
  {2020})}\BibitemShut {NoStop}%
\bibitem [{\citenamefont {Sneyd}\ \emph {et~al.}(2021)\citenamefont {Sneyd},
  \citenamefont {Fukui}, \citenamefont {Paleček}, \citenamefont {Prodhan},
  \citenamefont {Wagner}, \citenamefont {Zhang}, \citenamefont {Sung},
  \citenamefont {Collins}, \citenamefont {Slater}, \citenamefont
  {Andaji-Garmaroudi}, \citenamefont {MacFarlane}, \citenamefont
  {Garcia-Hernandez}, \citenamefont {Wang}, \citenamefont {Whittell},
  \citenamefont {Hodgkiss}, \citenamefont {Chen}, \citenamefont {Beljonne},
  \citenamefont {Manners}, \citenamefont {Friend},\ and\ \citenamefont
  {Rao}}]{Sneyd2021sciadv}%
  \BibitemOpen
  \bibfield  {author} {\bibinfo {author} {\bibfnamefont {A.~J.}\ \bibnamefont
  {Sneyd}}, \bibinfo {author} {\bibfnamefont {T.}~\bibnamefont {Fukui}},
  \bibinfo {author} {\bibfnamefont {D.}~\bibnamefont {Paleček}}, \bibinfo
  {author} {\bibfnamefont {S.}~\bibnamefont {Prodhan}}, \bibinfo {author}
  {\bibfnamefont {I.}~\bibnamefont {Wagner}}, \bibinfo {author} {\bibfnamefont
  {Y.}~\bibnamefont {Zhang}}, \bibinfo {author} {\bibfnamefont
  {J.}~\bibnamefont {Sung}}, \bibinfo {author} {\bibfnamefont {S.~M.}\
  \bibnamefont {Collins}}, \bibinfo {author} {\bibfnamefont {T.~J.~A.}\
  \bibnamefont {Slater}}, \bibinfo {author} {\bibfnamefont {Z.}~\bibnamefont
  {Andaji-Garmaroudi}}, \bibinfo {author} {\bibfnamefont {L.~R.}\ \bibnamefont
  {MacFarlane}}, \bibinfo {author} {\bibfnamefont {J.~D.}\ \bibnamefont
  {Garcia-Hernandez}}, \bibinfo {author} {\bibfnamefont {L.}~\bibnamefont
  {Wang}}, \bibinfo {author} {\bibfnamefont {G.~R.}\ \bibnamefont {Whittell}},
  \bibinfo {author} {\bibfnamefont {J.~M.}\ \bibnamefont {Hodgkiss}}, \bibinfo
  {author} {\bibfnamefont {K.}~\bibnamefont {Chen}}, \bibinfo {author}
  {\bibfnamefont {D.}~\bibnamefont {Beljonne}}, \bibinfo {author}
  {\bibfnamefont {I.}~\bibnamefont {Manners}}, \bibinfo {author} {\bibfnamefont
  {R.~H.}\ \bibnamefont {Friend}}, \ and\ \bibinfo {author} {\bibfnamefont
  {A.}~\bibnamefont {Rao}},\ }\bibfield  {title} {\enquote {\bibinfo {title}
  {Efficient energy transport in an organic semiconductor mediated by transient
  exciton delocalization},}\ }\href {\doibase 10.1126/sciadv.abh4232}
  {\bibfield  {journal} {\bibinfo  {journal} {Sci. Adv.}\ }\textbf {\bibinfo
  {volume} {7}},\ \bibinfo {pages} {eabh4232} (\bibinfo {year}
  {2021})}\BibitemShut {NoStop}%
\bibitem [{\citenamefont {Roosta}\ \emph {et~al.}(2022)\citenamefont {Roosta},
  \citenamefont {Ghalami}, \citenamefont {Elstner},\ and\ \citenamefont
  {Xie}}]{Roosta2022hopping}%
  \BibitemOpen
  \bibfield  {author} {\bibinfo {author} {\bibfnamefont {S.}~\bibnamefont
  {Roosta}}, \bibinfo {author} {\bibfnamefont {F.}~\bibnamefont {Ghalami}},
  \bibinfo {author} {\bibfnamefont {M.}~\bibnamefont {Elstner}}, \ and\
  \bibinfo {author} {\bibfnamefont {W.}~\bibnamefont {Xie}},\ }\bibfield
  {title} {\enquote {\bibinfo {title} {Efficient surface hopping approach for
  modeling charge transport in organic semiconductors},}\ }\href {\doibase
  10.1021/acs.jctc.1c00944} {\bibfield  {journal} {\bibinfo  {journal} {J.
  Chem. Theory Comput.}\ }\textbf {\bibinfo {volume} {18}},\ \bibinfo {pages}
  {1264--1274} (\bibinfo {year} {2022})}\BibitemShut {NoStop}%
\bibitem [{\citenamefont {Peng}\ \emph {et~al.}(2022)\citenamefont {Peng},
  \citenamefont {Brey}, \citenamefont {Giannini}, \citenamefont
  {Dell’Angelo}, \citenamefont {Burghardt},\ and\ \citenamefont
  {Blumberger}}]{Peng2022exciton}%
  \BibitemOpen
  \bibfield  {author} {\bibinfo {author} {\bibfnamefont {W.-T.}\ \bibnamefont
  {Peng}}, \bibinfo {author} {\bibfnamefont {D.}~\bibnamefont {Brey}}, \bibinfo
  {author} {\bibfnamefont {S.}~\bibnamefont {Giannini}}, \bibinfo {author}
  {\bibfnamefont {D.}~\bibnamefont {Dell’Angelo}}, \bibinfo {author}
  {\bibfnamefont {I.}~\bibnamefont {Burghardt}}, \ and\ \bibinfo {author}
  {\bibfnamefont {J.}~\bibnamefont {Blumberger}},\ }\bibfield  {title}
  {\enquote {\bibinfo {title} {Exciton dissociation in a model organic
  interface: Excitonic state-based surface hopping versus multiconfigurational
  time-dependent {H}artree},}\ }\href {\doibase 10.1021/acs.jpclett.2c01928}
  {\bibfield  {journal} {\bibinfo  {journal} {J. Phys. Chem. Lett.}\ }\textbf
  {\bibinfo {volume} {13}},\ \bibinfo {pages} {7105--7112} (\bibinfo {year}
  {2022})}\BibitemShut {NoStop}%
\bibitem [{\citenamefont {Giannini}\ and\ \citenamefont
  {Blumberger}(2022)}]{Giannini2022review}%
  \BibitemOpen
  \bibfield  {author} {\bibinfo {author} {\bibfnamefont {S.}~\bibnamefont
  {Giannini}}\ and\ \bibinfo {author} {\bibfnamefont {J.}~\bibnamefont
  {Blumberger}},\ }\bibfield  {title} {\enquote {\bibinfo {title} {Charge
  transport in organic semiconductors: The perspective from nonadiabatic
  molecular dynamics},}\ }\href {\doibase 10.1021/acs.accounts.1c00675}
  {\bibfield  {journal} {\bibinfo  {journal} {Acc. Chem. Res.}\ }\textbf
  {\bibinfo {volume} {55}},\ \bibinfo {pages} {819--830} (\bibinfo {year}
  {2022})},\ \bibinfo {note} {pMID: 35196456}\BibitemShut {NoStop}%
\bibitem [{\citenamefont {Wang}\ \emph {et~al.}(2020)\citenamefont {Wang},
  \citenamefont {Qiu}, \citenamefont {Bai},\ and\ \citenamefont
  {Xu}}]{Wang2020review}%
  \BibitemOpen
  \bibfield  {author} {\bibinfo {author} {\bibfnamefont {L.}~\bibnamefont
  {Wang}}, \bibinfo {author} {\bibfnamefont {J.}~\bibnamefont {Qiu}}, \bibinfo
  {author} {\bibfnamefont {X.}~\bibnamefont {Bai}}, \ and\ \bibinfo {author}
  {\bibfnamefont {J.}~\bibnamefont {Xu}},\ }\bibfield  {title} {\enquote
  {\bibinfo {title} {Surface hopping methods for nonadiabatic dynamics in
  extended systems},}\ }\href {\doibase 10.1002/wcms.1435} {\bibfield
  {journal} {\bibinfo  {journal} {Wiley Interdiscip. Rev.: Comput. Mol. Sci.}\
  }\textbf {\bibinfo {volume} {10}},\ \bibinfo {pages} {e1435} (\bibinfo {year}
  {2020})}\BibitemShut {NoStop}%
\bibitem [{\citenamefont {Bai}, \citenamefont {Qiu},\ and\ \citenamefont
  {Wang}(2018)}]{Bai2018trivial}%
  \BibitemOpen
  \bibfield  {author} {\bibinfo {author} {\bibfnamefont {X.}~\bibnamefont
  {Bai}}, \bibinfo {author} {\bibfnamefont {J.}~\bibnamefont {Qiu}}, \ and\
  \bibinfo {author} {\bibfnamefont {L.}~\bibnamefont {Wang}},\ }\bibfield
  {title} {\enquote {\bibinfo {title} {{An efficient solution to the
  decoherence enhanced trivial crossing problem in surface hopping}},}\ }\href
  {\doibase 10.1063/1.5020693} {\bibfield  {journal} {\bibinfo  {journal} {J.
  Chem. Phys.}\ }\textbf {\bibinfo {volume} {148}},\ \bibinfo {pages} {104106}
  (\bibinfo {year} {2018})}\BibitemShut {NoStop}%
\bibitem [{\citenamefont {Carof}, \citenamefont {Giannini},\ and\ \citenamefont
  {Blumberger}(2019)}]{Carof2019mobility}%
  \BibitemOpen
  \bibfield  {author} {\bibinfo {author} {\bibfnamefont {A.}~\bibnamefont
  {Carof}}, \bibinfo {author} {\bibfnamefont {S.}~\bibnamefont {Giannini}}, \
  and\ \bibinfo {author} {\bibfnamefont {J.}~\bibnamefont {Blumberger}},\
  }\bibfield  {title} {\enquote {\bibinfo {title} {How to calculate charge
  mobility in molecular materials from surface hopping non-adiabatic molecular
  dynamics – beyond the hopping/band paradigm},}\ }\href {\doibase
  10.1039/C9CP04770K} {\bibfield  {journal} {\bibinfo  {journal} {Phys. Chem.
  Chem. Phys.}\ }\textbf {\bibinfo {volume} {21}},\ \bibinfo {pages}
  {26368--26386} (\bibinfo {year} {2019})}\BibitemShut {NoStop}%
\bibitem [{\citenamefont {Tuckerman}(2010)}]{Tuckerman2010book}%
  \BibitemOpen
  \bibfield  {author} {\bibinfo {author} {\bibfnamefont {M.~E.}\ \bibnamefont
  {Tuckerman}},\ }\href@noop {} {\emph {\bibinfo {title} {Statistical
  Mechanics: Theory and Molecular Simulation}}}\ (\bibinfo  {publisher} {Oxford
  University Press},\ \bibinfo {year} {2010})\BibitemShut {NoStop}%
\bibitem [{\citenamefont {Mannouch}\ and\ \citenamefont
  {Richardson}(2023)}]{Mannouch2023mash}%
  \BibitemOpen
  \bibfield  {author} {\bibinfo {author} {\bibfnamefont {J.~R.}\ \bibnamefont
  {Mannouch}}\ and\ \bibinfo {author} {\bibfnamefont {J.~O.}\ \bibnamefont
  {Richardson}},\ }\bibfield  {title} {\enquote {\bibinfo {title} {{A mapping
  approach to surface hopping}},}\ }\href {https://doi.org/10.1063/5.0139734}
  {\bibfield  {journal} {\bibinfo  {journal} {J. Chem. Phys.}\ }\textbf
  {\bibinfo {volume} {158}},\ \bibinfo {pages} {104111} (\bibinfo {year}
  {2023})}\BibitemShut {NoStop}%
\bibitem [{\citenamefont {Runeson}\ and\ \citenamefont
  {Manolopoulos}(2023)}]{Runeson2023mash}%
  \BibitemOpen
  \bibfield  {author} {\bibinfo {author} {\bibfnamefont {J.~E.}\ \bibnamefont
  {Runeson}}\ and\ \bibinfo {author} {\bibfnamefont {D.~E.}\ \bibnamefont
  {Manolopoulos}},\ }\bibfield  {title} {\enquote {\bibinfo {title} {{A
  multi-state mapping approach to surface hopping}},}\ }\href {\doibase
  10.1063/5.0158147} {\bibfield  {journal} {\bibinfo  {journal} {J. Chem.
  Phys.}\ }\textbf {\bibinfo {volume} {159}},\ \bibinfo {pages} {094115}
  (\bibinfo {year} {2023})}\BibitemShut {NoStop}%
\bibitem [{Note1()}]{Note1}%
  \BibitemOpen
  \bibinfo {note} {Note that we are referring here to the version of MASH
  described in Ref.~\protect \rev@citealpnum {Runeson2023mash}, not to the more
  sophisticated ``uncoupled spheres'' MASH method that was recently developed
  for applications to gas-phase photochemistry (Refs.~\protect \rev@citealpnum
  {Lawrence2024sizeconsistent} and \protect \rev@citealpnum
  {Lawrence2024cyclobutanone}).}\BibitemShut {Stop}%
\bibitem [{\citenamefont {Lawrence}, \citenamefont {Mannouch},\ and\
  \citenamefont {Richardson}(2024{\natexlab{a}})}]{Lawrence2024sizeconsistent}%
  \BibitemOpen
  \bibfield  {author} {\bibinfo {author} {\bibfnamefont {J.~E.}\ \bibnamefont
  {Lawrence}}, \bibinfo {author} {\bibfnamefont {J.~R.}\ \bibnamefont
  {Mannouch}}, \ and\ \bibinfo {author} {\bibfnamefont {J.~O.}\ \bibnamefont
  {Richardson}},\ }\bibfield  {title} {\enquote {\bibinfo {title} {{A
  size-consistent multi-state mapping approach to surface hopping}},}\ }\href
  {\doibase 10.1063/5.0208575} {\bibfield  {journal} {\bibinfo  {journal} {J.
  Chem. Phys.}\ }\textbf {\bibinfo {volume} {160}},\ \bibinfo {pages} {244112}
  (\bibinfo {year} {2024}{\natexlab{a}})}\BibitemShut {NoStop}%
\bibitem [{\citenamefont {Lawrence}\ \emph {et~al.}(2024)\citenamefont
  {Lawrence}, \citenamefont {Ansari}, \citenamefont {Mannouch}, \citenamefont
  {Manae}, \citenamefont {Asnaashari}, \citenamefont {Kelly},\ and\
  \citenamefont {Richardson}}]{Lawrence2024cyclobutanone}%
  \BibitemOpen
  \bibfield  {author} {\bibinfo {author} {\bibfnamefont {J.~E.}\ \bibnamefont
  {Lawrence}}, \bibinfo {author} {\bibfnamefont {I.~M.}\ \bibnamefont
  {Ansari}}, \bibinfo {author} {\bibfnamefont {J.~R.}\ \bibnamefont
  {Mannouch}}, \bibinfo {author} {\bibfnamefont {M.~A.}\ \bibnamefont {Manae}},
  \bibinfo {author} {\bibfnamefont {K.}~\bibnamefont {Asnaashari}}, \bibinfo
  {author} {\bibfnamefont {A.}~\bibnamefont {Kelly}}, \ and\ \bibinfo {author}
  {\bibfnamefont {J.~O.}\ \bibnamefont {Richardson}},\ }\bibfield  {title}
  {\enquote {\bibinfo {title} {{A MASH simulation of the photoexcited dynamics
  of cyclobutanone}},}\ }\href {\doibase 10.1063/5.0203695} {\bibfield
  {journal} {\bibinfo  {journal} {J. Chem. Phys.}\ }\textbf {\bibinfo {volume}
  {160}},\ \bibinfo {pages} {174306} (\bibinfo {year} {2024})}\BibitemShut
  {NoStop}%
\bibitem [{\citenamefont {Amati}, \citenamefont {Mannouch},\ and\ \citenamefont
  {Richardson}(2023)}]{Amati2023thermalization}%
  \BibitemOpen
  \bibfield  {author} {\bibinfo {author} {\bibfnamefont {G.}~\bibnamefont
  {Amati}}, \bibinfo {author} {\bibfnamefont {J.~R.}\ \bibnamefont {Mannouch}},
  \ and\ \bibinfo {author} {\bibfnamefont {J.~O.}\ \bibnamefont {Richardson}},\
  }\bibfield  {title} {\enquote {\bibinfo {title} {{Detailed balance in mixed
  quantum–classical mapping approaches}},}\ }\href {\doibase
  10.1063/5.0176291} {\bibfield  {journal} {\bibinfo  {journal} {J. Chem.
  Phys.}\ }\textbf {\bibinfo {volume} {159}},\ \bibinfo {pages} {214114}
  (\bibinfo {year} {2023})}\BibitemShut {NoStop}%
\bibitem [{\citenamefont {Runeson}, \citenamefont {Fay},\ and\ \citenamefont
  {Manolopoulos}(2024)}]{Runeson2023pccp}%
  \BibitemOpen
  \bibfield  {author} {\bibinfo {author} {\bibfnamefont {J.~E.}\ \bibnamefont
  {Runeson}}, \bibinfo {author} {\bibfnamefont {T.~P.}\ \bibnamefont {Fay}}, \
  and\ \bibinfo {author} {\bibfnamefont {D.~E.}\ \bibnamefont {Manolopoulos}},\
  }\bibfield  {title} {\enquote {\bibinfo {title} {Exciton dynamics from the
  mapping approach to surface hopping: {C}omparison with {F}\"orster and
  {R}edfield theories},}\ }\href {\doibase 10.1039/D3CP05926J} {\bibfield
  {journal} {\bibinfo  {journal} {Phys. Chem. Chem. Phys.}\ }\textbf {\bibinfo
  {volume} {26}},\ \bibinfo {pages} {4929--4938} (\bibinfo {year}
  {2024})}\BibitemShut {NoStop}%
\bibitem [{\citenamefont {Lawrence}, \citenamefont {Mannouch},\ and\
  \citenamefont {Richardson}(2024{\natexlab{b}})}]{Lawrence2023mash}%
  \BibitemOpen
  \bibfield  {author} {\bibinfo {author} {\bibfnamefont {J.~E.}\ \bibnamefont
  {Lawrence}}, \bibinfo {author} {\bibfnamefont {J.~R.}\ \bibnamefont
  {Mannouch}}, \ and\ \bibinfo {author} {\bibfnamefont {J.~O.}\ \bibnamefont
  {Richardson}},\ }\bibfield  {title} {\enquote {\bibinfo {title} {Recovering
  marcus theory rates and beyond without the need for decoherence corrections:
  The mapping approach to surface hopping},}\ }\href {\doibase
  10.1021/acs.jpclett.3c03197} {\bibfield  {journal} {\bibinfo  {journal} {J.
  Phys. Chem. Lett.}\ }\textbf {\bibinfo {volume} {15}},\ \bibinfo {pages}
  {707--716} (\bibinfo {year} {2024}{\natexlab{b}})}\BibitemShut {NoStop}%
\bibitem [{\citenamefont {Wang}\ \emph {et~al.}(2011)\citenamefont {Wang},
  \citenamefont {Beljonne}, \citenamefont {Chen},\ and\ \citenamefont
  {Shi}}]{Wang2011charge}%
  \BibitemOpen
  \bibfield  {author} {\bibinfo {author} {\bibfnamefont {L.}~\bibnamefont
  {Wang}}, \bibinfo {author} {\bibfnamefont {D.}~\bibnamefont {Beljonne}},
  \bibinfo {author} {\bibfnamefont {L.}~\bibnamefont {Chen}}, \ and\ \bibinfo
  {author} {\bibfnamefont {Q.}~\bibnamefont {Shi}},\ }\bibfield  {title}
  {\enquote {\bibinfo {title} {{Mixed quantum-classical simulations of charge
  transport in organic materials: Numerical benchmark of the
  Su-Schrieffer-Heeger model}},}\ }\href {\doibase 10.1063/1.3604561}
  {\bibfield  {journal} {\bibinfo  {journal} {J. Chem. Phys.}\ }\textbf
  {\bibinfo {volume} {134}},\ \bibinfo {pages} {244116} (\bibinfo {year}
  {2011})}\BibitemShut {NoStop}%
\bibitem [{\citenamefont {Fetherolf}, \citenamefont
  {Gole\ifmmode~\check{z}\else \v{z}\fi{}},\ and\ \citenamefont
  {Berkelbach}(2020)}]{Fetherolf2020prx}%
  \BibitemOpen
  \bibfield  {author} {\bibinfo {author} {\bibfnamefont {J.~H.}\ \bibnamefont
  {Fetherolf}}, \bibinfo {author} {\bibfnamefont {D.}~\bibnamefont
  {Gole\ifmmode~\check{z}\else \v{z}\fi{}}}, \ and\ \bibinfo {author}
  {\bibfnamefont {T.~C.}\ \bibnamefont {Berkelbach}},\ }\bibfield  {title}
  {\enquote {\bibinfo {title} {A unification of the holstein polaron and
  dynamic disorder pictures of charge transport in organic crystals},}\ }\href
  {\doibase 10.1103/PhysRevX.10.021062} {\bibfield  {journal} {\bibinfo
  {journal} {Phys. Rev. X}\ }\textbf {\bibinfo {volume} {10}},\ \bibinfo
  {pages} {021062} (\bibinfo {year} {2020})}\BibitemShut {NoStop}%
\bibitem [{\citenamefont {Fetherolf}, \citenamefont {Shih},\ and\ \citenamefont
  {Berkelbach}(2023)}]{Fetherolf2023prb}%
  \BibitemOpen
  \bibfield  {author} {\bibinfo {author} {\bibfnamefont {J.~H.}\ \bibnamefont
  {Fetherolf}}, \bibinfo {author} {\bibfnamefont {P.}~\bibnamefont {Shih}}, \
  and\ \bibinfo {author} {\bibfnamefont {T.~C.}\ \bibnamefont {Berkelbach}},\
  }\bibfield  {title} {\enquote {\bibinfo {title} {Conductivity of an electron
  coupled to anharmonic phonons: Quantum-classical simulations and comparison
  of approximations},}\ }\href {\doibase 10.1103/PhysRevB.107.064304}
  {\bibfield  {journal} {\bibinfo  {journal} {Phys. Rev. B}\ }\textbf {\bibinfo
  {volume} {107}},\ \bibinfo {pages} {064304} (\bibinfo {year}
  {2023})}\BibitemShut {NoStop}%
\bibitem [{\citenamefont {Cataudella}, \citenamefont {De~Filippis},\ and\
  \citenamefont {Perroni}(2011)}]{Cataudella2011prb}%
  \BibitemOpen
  \bibfield  {author} {\bibinfo {author} {\bibfnamefont {V.}~\bibnamefont
  {Cataudella}}, \bibinfo {author} {\bibfnamefont {G.}~\bibnamefont
  {De~Filippis}}, \ and\ \bibinfo {author} {\bibfnamefont {C.~A.}\ \bibnamefont
  {Perroni}},\ }\bibfield  {title} {\enquote {\bibinfo {title} {{Transport
  properties and optical conductivity of the adiabatic Su-Schrieffer-Heeger
  model: A showcase study for rubrene-based field effect transistors}},}\
  }\href {\doibase 10.1103/PhysRevB.83.165203} {\bibfield  {journal} {\bibinfo
  {journal} {Phys. Rev. B}\ }\textbf {\bibinfo {volume} {83}},\ \bibinfo
  {pages} {165203} (\bibinfo {year} {2011})}\BibitemShut {NoStop}%
\bibitem [{\citenamefont {Zhong}\ and\ \citenamefont
  {Zhao}(2011)}]{Zhong2011sse}%
  \BibitemOpen
  \bibfield  {author} {\bibinfo {author} {\bibfnamefont {X.}~\bibnamefont
  {Zhong}}\ and\ \bibinfo {author} {\bibfnamefont {Y.}~\bibnamefont {Zhao}},\
  }\bibfield  {title} {\enquote {\bibinfo {title} {{Charge carrier dynamics in
  phonon-induced fluctuation systems from time-dependent wavepacket diffusion
  approach}},}\ }\href {\doibase 10.1063/1.3644965} {\bibfield  {journal}
  {\bibinfo  {journal} {J. Chem. Phys.}\ }\textbf {\bibinfo {volume} {135}},\
  \bibinfo {pages} {134110} (\bibinfo {year} {2011})}\BibitemShut {NoStop}%
\bibitem [{\citenamefont {Troisi}(2007)}]{Troisi2007ehrenfest}%
  \BibitemOpen
  \bibfield  {author} {\bibinfo {author} {\bibfnamefont {A.}~\bibnamefont
  {Troisi}},\ }\bibfield  {title} {\enquote {\bibinfo {title} {Prediction of
  the absolute charge mobility of molecular semiconductors: the case of
  rubrene},}\ }\href {\doibase https://doi.org/10.1002/adma.200700550}
  {\bibfield  {journal} {\bibinfo  {journal} {Adv. Mater.}\ }\textbf {\bibinfo
  {volume} {19}},\ \bibinfo {pages} {2000--2004} (\bibinfo {year}
  {2007})}\BibitemShut {NoStop}%
\bibitem [{\citenamefont {Poole}\ \emph {et~al.}(2016)\citenamefont {Poole},
  \citenamefont {Damry}, \citenamefont {Tozer},\ and\ \citenamefont
  {Barford}}]{Poole2016mobility}%
  \BibitemOpen
  \bibfield  {author} {\bibinfo {author} {\bibfnamefont {J.~E.}\ \bibnamefont
  {Poole}}, \bibinfo {author} {\bibfnamefont {D.~A.}\ \bibnamefont {Damry}},
  \bibinfo {author} {\bibfnamefont {O.~R.}\ \bibnamefont {Tozer}}, \ and\
  \bibinfo {author} {\bibfnamefont {W.}~\bibnamefont {Barford}},\ }\bibfield
  {title} {\enquote {\bibinfo {title} {Charge mobility induced by {B}rownian
  fluctuations in $\pi$-conjugated polymers in solution},}\ }\href {\doibase
  10.1039/C5CP06842H} {\bibfield  {journal} {\bibinfo  {journal} {Phys. Chem.
  Chem. Phys.}\ }\textbf {\bibinfo {volume} {18}},\ \bibinfo {pages}
  {2574--2579} (\bibinfo {year} {2016})}\BibitemShut {NoStop}%
\bibitem [{\citenamefont {Hegger}, \citenamefont {Binder},\ and\ \citenamefont
  {Burghardt}(2020)}]{Hegger2020jctc}%
  \BibitemOpen
  \bibfield  {author} {\bibinfo {author} {\bibfnamefont {R.}~\bibnamefont
  {Hegger}}, \bibinfo {author} {\bibfnamefont {R.}~\bibnamefont {Binder}}, \
  and\ \bibinfo {author} {\bibfnamefont {I.}~\bibnamefont {Burghardt}},\
  }\bibfield  {title} {\enquote {\bibinfo {title} {First-principles quantum and
  quantum-classical simulations of exciton diffusion in semiconducting polymer
  chains at finite temperature},}\ }\href {\doibase 10.1021/acs.jctc.0c00351}
  {\bibfield  {journal} {\bibinfo  {journal} {J. Chem. Theory Comput.}\
  }\textbf {\bibinfo {volume} {16}},\ \bibinfo {pages} {5441--5455} (\bibinfo
  {year} {2020})}\BibitemShut {NoStop}%
\bibitem [{\citenamefont {Berencei}, \citenamefont {Barford},\ and\
  \citenamefont {Clark}(2022)}]{Berencei2022}%
  \BibitemOpen
  \bibfield  {author} {\bibinfo {author} {\bibfnamefont {L.}~\bibnamefont
  {Berencei}}, \bibinfo {author} {\bibfnamefont {W.}~\bibnamefont {Barford}}, \
  and\ \bibinfo {author} {\bibfnamefont {S.~R.}\ \bibnamefont {Clark}},\
  }\bibfield  {title} {\enquote {\bibinfo {title} {Thermally driven polaron
  transport in conjugated polymers},}\ }\href {\doibase
  10.1103/PhysRevB.105.014303} {\bibfield  {journal} {\bibinfo  {journal}
  {Phys. Rev. B}\ }\textbf {\bibinfo {volume} {105}},\ \bibinfo {pages}
  {014303} (\bibinfo {year} {2022})}\BibitemShut {NoStop}%
\bibitem [{\citenamefont {Parandekar}\ and\ \citenamefont
  {Tully}(2006)}]{parandekar2006ehrenfest}%
  \BibitemOpen
  \bibfield  {author} {\bibinfo {author} {\bibfnamefont {P.~V.}\ \bibnamefont
  {Parandekar}}\ and\ \bibinfo {author} {\bibfnamefont {J.~C.}\ \bibnamefont
  {Tully}},\ }\bibfield  {title} {\enquote {\bibinfo {title} {Detailed balance
  in {E}hrenfest mixed quantum-classical dynamics},}\ }\href {\doibase
  10.1021/ct050213k} {\bibfield  {journal} {\bibinfo  {journal} {J. Chem.
  Theory Comput.}\ }\textbf {\bibinfo {volume} {2}},\ \bibinfo {pages}
  {229--235} (\bibinfo {year} {2006})}\BibitemShut {NoStop}%
\bibitem [{\citenamefont {Runeson}\ \emph {et~al.}(2022)\citenamefont
  {Runeson}, \citenamefont {Lawrence}, \citenamefont {Mannouch},\ and\
  \citenamefont {Richardson}}]{runeson2022fmo}%
  \BibitemOpen
  \bibfield  {author} {\bibinfo {author} {\bibfnamefont {J.~E.}\ \bibnamefont
  {Runeson}}, \bibinfo {author} {\bibfnamefont {J.~E.}\ \bibnamefont
  {Lawrence}}, \bibinfo {author} {\bibfnamefont {J.~R.}\ \bibnamefont
  {Mannouch}}, \ and\ \bibinfo {author} {\bibfnamefont {J.~O.}\ \bibnamefont
  {Richardson}},\ }\bibfield  {title} {\enquote {\bibinfo {title} {Explaining
  the efficiency of photosynthesis: Quantum uncertainty or classical
  vibrations?}}\ }\href {\doibase 10.1021/acs.jpclett.2c00538} {\bibfield
  {journal} {\bibinfo  {journal} {J. Phys. Chem. Lett.}\ }\textbf {\bibinfo
  {volume} {13}},\ \bibinfo {pages} {3392--3399} (\bibinfo {year}
  {2022})}\BibitemShut {NoStop}%
\bibitem [{\citenamefont {Schmidt}, \citenamefont {Parandekar},\ and\
  \citenamefont {Tully}(2008)}]{schmidt2008SH}%
  \BibitemOpen
  \bibfield  {author} {\bibinfo {author} {\bibfnamefont {J.~R.}\ \bibnamefont
  {Schmidt}}, \bibinfo {author} {\bibfnamefont {P.~V.}\ \bibnamefont
  {Parandekar}}, \ and\ \bibinfo {author} {\bibfnamefont {J.~C.}\ \bibnamefont
  {Tully}},\ }\bibfield  {title} {\enquote {\bibinfo {title} {Mixed
  quantum-classical equilibrium: Surface hopping},}\ }\href {\doibase
  10.1063/1.2955564} {\bibfield  {journal} {\bibinfo  {journal} {J.~Chem.
  Phys.}\ }\textbf {\bibinfo {volume} {129}},\ \bibinfo {pages} {044104}
  (\bibinfo {year} {2008})}\BibitemShut {NoStop}%
\bibitem [{\citenamefont {Podzorov}\ \emph {et~al.}(2005)\citenamefont
  {Podzorov}, \citenamefont {Menard}, \citenamefont {Rogers},\ and\
  \citenamefont {Gershenson}}]{Podzorov2005}%
  \BibitemOpen
  \bibfield  {author} {\bibinfo {author} {\bibfnamefont {V.}~\bibnamefont
  {Podzorov}}, \bibinfo {author} {\bibfnamefont {E.}~\bibnamefont {Menard}},
  \bibinfo {author} {\bibfnamefont {J.~A.}\ \bibnamefont {Rogers}}, \ and\
  \bibinfo {author} {\bibfnamefont {M.~E.}\ \bibnamefont {Gershenson}},\
  }\bibfield  {title} {\enquote {\bibinfo {title} {Hall effect in the
  accumulation layers on the surface of organic semiconductors},}\ }\href
  {\doibase 10.1103/PhysRevLett.95.226601} {\bibfield  {journal} {\bibinfo
  {journal} {Phys. Rev. Lett.}\ }\textbf {\bibinfo {volume} {95}},\ \bibinfo
  {pages} {226601} (\bibinfo {year} {2005})}\BibitemShut {NoStop}%
\bibitem [{\citenamefont {Fischer}\ \emph {et~al.}(2006)\citenamefont
  {Fischer}, \citenamefont {Dressel}, \citenamefont {Gompf}, \citenamefont
  {Tripathi},\ and\ \citenamefont {Pflaum}}]{Fischer2006infrared}%
  \BibitemOpen
  \bibfield  {author} {\bibinfo {author} {\bibfnamefont {M.}~\bibnamefont
  {Fischer}}, \bibinfo {author} {\bibfnamefont {M.}~\bibnamefont {Dressel}},
  \bibinfo {author} {\bibfnamefont {B.}~\bibnamefont {Gompf}}, \bibinfo
  {author} {\bibfnamefont {A.~K.}\ \bibnamefont {Tripathi}}, \ and\ \bibinfo
  {author} {\bibfnamefont {J.}~\bibnamefont {Pflaum}},\ }\bibfield  {title}
  {\enquote {\bibinfo {title} {{Infrared spectroscopy on the charge
  accumulation layer in rubrene single crystals}},}\ }\href {\doibase
  10.1063/1.2370743} {\bibfield  {journal} {\bibinfo  {journal} {Appl. Phys.
  Lett.}\ }\textbf {\bibinfo {volume} {89}},\ \bibinfo {pages} {182103}
  (\bibinfo {year} {2006})}\BibitemShut {NoStop}%
\bibitem [{\citenamefont {Li}\ \emph {et~al.}(2007)\citenamefont {Li},
  \citenamefont {Podzorov}, \citenamefont {Sai}, \citenamefont {Martin},
  \citenamefont {Gershenson}, \citenamefont {Di~Ventra},\ and\ \citenamefont
  {Basov}}]{LiPodzorov2007}%
  \BibitemOpen
  \bibfield  {author} {\bibinfo {author} {\bibfnamefont {Z.~Q.}\ \bibnamefont
  {Li}}, \bibinfo {author} {\bibfnamefont {V.}~\bibnamefont {Podzorov}},
  \bibinfo {author} {\bibfnamefont {N.}~\bibnamefont {Sai}}, \bibinfo {author}
  {\bibfnamefont {M.~C.}\ \bibnamefont {Martin}}, \bibinfo {author}
  {\bibfnamefont {M.~E.}\ \bibnamefont {Gershenson}}, \bibinfo {author}
  {\bibfnamefont {M.}~\bibnamefont {Di~Ventra}}, \ and\ \bibinfo {author}
  {\bibfnamefont {D.~N.}\ \bibnamefont {Basov}},\ }\bibfield  {title} {\enquote
  {\bibinfo {title} {Light quasiparticles dominate electronic transport in
  molecular crystal field-effect transistors},}\ }\href {\doibase
  10.1103/PhysRevLett.99.016403} {\bibfield  {journal} {\bibinfo  {journal}
  {Phys. Rev. Lett.}\ }\textbf {\bibinfo {volume} {99}},\ \bibinfo {pages}
  {016403} (\bibinfo {year} {2007})}\BibitemShut {NoStop}%
\bibitem [{\citenamefont {Fratini}, \citenamefont {Ciuchi},\ and\ \citenamefont
  {Mayou}(2014)}]{Fratini2014phenomenological}%
  \BibitemOpen
  \bibfield  {author} {\bibinfo {author} {\bibfnamefont {S.}~\bibnamefont
  {Fratini}}, \bibinfo {author} {\bibfnamefont {S.}~\bibnamefont {Ciuchi}}, \
  and\ \bibinfo {author} {\bibfnamefont {D.}~\bibnamefont {Mayou}},\ }\bibfield
   {title} {\enquote {\bibinfo {title} {Phenomenological model for charge
  dynamics and optical response of disordered systems: Application to organic
  semiconductors},}\ }\href {\doibase 10.1103/PhysRevB.89.235201} {\bibfield
  {journal} {\bibinfo  {journal} {Phys. Rev. B}\ }\textbf {\bibinfo {volume}
  {89}},\ \bibinfo {pages} {235201} (\bibinfo {year} {2014})}\BibitemShut
  {NoStop}%
\bibitem [{\citenamefont {Nitzan}(2006)}]{nitzan2006book}%
  \BibitemOpen
  \bibfield  {author} {\bibinfo {author} {\bibfnamefont {A.}~\bibnamefont
  {Nitzan}},\ }\href@noop {} {\emph {\bibinfo {title} {Chemical Dynamics in
  Condensed Phases: Relaxation, Transfer, and Reactions in Condensed Molecular
  Systems}}}\ (\bibinfo  {publisher} {Oxford University Press},\ \bibinfo
  {address} {Oxford},\ \bibinfo {year} {2006})\BibitemShut {NoStop}%
\bibitem [{\citenamefont {Troisi}\ and\ \citenamefont
  {Cheung}(2009)}]{Troisi2009disorder}%
  \BibitemOpen
  \bibfield  {author} {\bibinfo {author} {\bibfnamefont {A.}~\bibnamefont
  {Troisi}}\ and\ \bibinfo {author} {\bibfnamefont {D.~L.}\ \bibnamefont
  {Cheung}},\ }\bibfield  {title} {\enquote {\bibinfo {title} {{Transition from
  dynamic to static disorder in one-dimensional organic semiconductors}},}\
  }\href {\doibase 10.1063/1.3167406} {\bibfield  {journal} {\bibinfo
  {journal} {J. Chem. Phys.}\ }\textbf {\bibinfo {volume} {131}},\ \bibinfo
  {pages} {014703} (\bibinfo {year} {2009})}\BibitemShut {NoStop}%
\bibitem [{\citenamefont {Dettmann}\ \emph {et~al.}(2023)\citenamefont
  {Dettmann}, \citenamefont {Cavalcante}, \citenamefont {Magdaleno},\ and\
  \citenamefont {Moulé}}]{Dettmann2023killer}%
  \BibitemOpen
  \bibfield  {author} {\bibinfo {author} {\bibfnamefont {M.~A.}\ \bibnamefont
  {Dettmann}}, \bibinfo {author} {\bibfnamefont {L.~S.~R.}\ \bibnamefont
  {Cavalcante}}, \bibinfo {author} {\bibfnamefont {C.~A.}\ \bibnamefont
  {Magdaleno}}, \ and\ \bibinfo {author} {\bibfnamefont {A.~J.}\ \bibnamefont
  {Moulé}},\ }\bibfield  {title} {\enquote {\bibinfo {title} {Catching the
  killer: Dynamic disorder design rules for small-molecule organic
  semiconductors},}\ }\href {\doibase https://doi.org/10.1002/adfm.202213370}
  {\bibfield  {journal} {\bibinfo  {journal} {Adv. Funct. Mater.}\ }\textbf
  {\bibinfo {volume} {33}},\ \bibinfo {pages} {2213370} (\bibinfo {year}
  {2023})}\BibitemShut {NoStop}%
\bibitem [{\citenamefont {Knepp}\ and\ \citenamefont
  {Fredin}(2024)}]{Knepp2024boltzmann}%
  \BibitemOpen
  \bibfield  {author} {\bibinfo {author} {\bibfnamefont {Z.~J.}\ \bibnamefont
  {Knepp}}\ and\ \bibinfo {author} {\bibfnamefont {L.~A.}\ \bibnamefont
  {Fredin}},\ }\bibfield  {title} {\enquote {\bibinfo {title}
  {Finite-displacement boltzmann transport theory reveals the detrimental
  effects of high-frequency normal modes on mobility},}\ }\href {\doibase
  10.1103/PhysRevB.109.094307} {\bibfield  {journal} {\bibinfo  {journal}
  {Phys. Rev. B}\ }\textbf {\bibinfo {volume} {109}},\ \bibinfo {pages}
  {094307} (\bibinfo {year} {2024})}\BibitemShut {NoStop}%
\bibitem [{\citenamefont {Dunn}, \citenamefont {Tempelaar},\ and\ \citenamefont
  {Reichman}(2019)}]{Dunn2019heom}%
  \BibitemOpen
  \bibfield  {author} {\bibinfo {author} {\bibfnamefont {I.~S.}\ \bibnamefont
  {Dunn}}, \bibinfo {author} {\bibfnamefont {R.}~\bibnamefont {Tempelaar}}, \
  and\ \bibinfo {author} {\bibfnamefont {D.~R.}\ \bibnamefont {Reichman}},\
  }\bibfield  {title} {\enquote {\bibinfo {title} {{Removing instabilities in
  the hierarchical equations of motion: Exact and approximate projection
  approaches}},}\ }\href {\doibase 10.1063/1.5092616} {\bibfield  {journal}
  {\bibinfo  {journal} {J. Chem. Phys.}\ }\textbf {\bibinfo {volume} {150}},\
  \bibinfo {pages} {184109} (\bibinfo {year} {2019})}\BibitemShut {NoStop}%
\bibitem [{\citenamefont {Jankovi\'{c}}\ and\ \citenamefont
  {Vukmirovi\'{c}}(2022)}]{Jankovic2022prb}%
  \BibitemOpen
  \bibfield  {author} {\bibinfo {author} {\bibfnamefont {V.}~\bibnamefont
  {Jankovi\'{c}}}\ and\ \bibinfo {author} {\bibfnamefont {N.}~\bibnamefont
  {Vukmirovi\'{c}}},\ }\bibfield  {title} {\enquote {\bibinfo {title} {Spectral
  and thermodynamic properties of the holstein polaron: Hierarchical equations
  of motion approach},}\ }\href {\doibase 10.1103/PhysRevB.105.054311}
  {\bibfield  {journal} {\bibinfo  {journal} {Phys. Rev. B}\ }\textbf {\bibinfo
  {volume} {105}},\ \bibinfo {pages} {054311} (\bibinfo {year}
  {2022})}\BibitemShut {NoStop}%
\bibitem [{Note2()}]{Note2}%
  \BibitemOpen
  \bibinfo {note} {One way to see this is to note that the imaginary time
  correlation can be written as an integral over the real time correlation
  function with a kernel that decays as $e^{-2\pi t/\beta \hbar }$. See
  Eq.~(22) of Ref.~\protect \rev@citealpnum {Habershon2007meac}}\BibitemShut
  {NoStop}%
\bibitem [{\citenamefont {Habershon}, \citenamefont {Braams},\ and\
  \citenamefont {Manolopoulos}(2007)}]{Habershon2007meac}%
  \BibitemOpen
  \bibfield  {author} {\bibinfo {author} {\bibfnamefont {S.}~\bibnamefont
  {Habershon}}, \bibinfo {author} {\bibfnamefont {B.~J.}\ \bibnamefont
  {Braams}}, \ and\ \bibinfo {author} {\bibfnamefont {D.~E.}\ \bibnamefont
  {Manolopoulos}},\ }\bibfield  {title} {\enquote {\bibinfo {title} {{Quantum
  mechanical correlation functions, maximum entropy analytic continuation, and
  ring polymer molecular dynamics}},}\ }\href {\doibase 10.1063/1.2786451}
  {\bibfield  {journal} {\bibinfo  {journal} {J. Chem. Phys.}\ }\textbf
  {\bibinfo {volume} {127}},\ \bibinfo {pages} {174108} (\bibinfo {year}
  {2007})}\BibitemShut {NoStop}%
\end{thebibliography}%

\end{document}

\title{Charge transport in organic semiconductors from the mapping approach to surface hopping: Supplementary information}
\author{Johan E. Runeson}
\email{johan.runeson@chem.ox.ac.uk}
\affiliation{Department of Chemistry, University of Oxford, Physical and Theoretical Chemistry Laboratory, South Parks Road, Oxford, OX1 3QZ, UK}
\author{Thomas J. G. Drayton}
\affiliation{Department of Chemistry, University of Oxford, Physical and Theoretical Chemistry Laboratory, South Parks Road, Oxford, OX1 3QZ, UK}
\author{David E. Manolopoulos}
\affiliation{Department of Chemistry, University of Oxford, Physical and Theoretical Chemistry Laboratory, South Parks Road, Oxford, OX1 3QZ, UK}

\begin{abstract}
Here we discuss finite size effects in Ehrenfest and CPA calculations of charge transport in the rubrene model and present the convergence tests we have performed for these methods and MASH as a function of the number of lattice sites $N$. The situation we consider is the same as in Fig.~1 of our paper, which shows the time-dependent diffusivity $D(t)$ at 300 K.
\end{abstract}

\maketitle

\subsection{Ehrenfest dynamics}

\begin{figure}[h]
    \centering
    \includegraphics[scale=0.95]{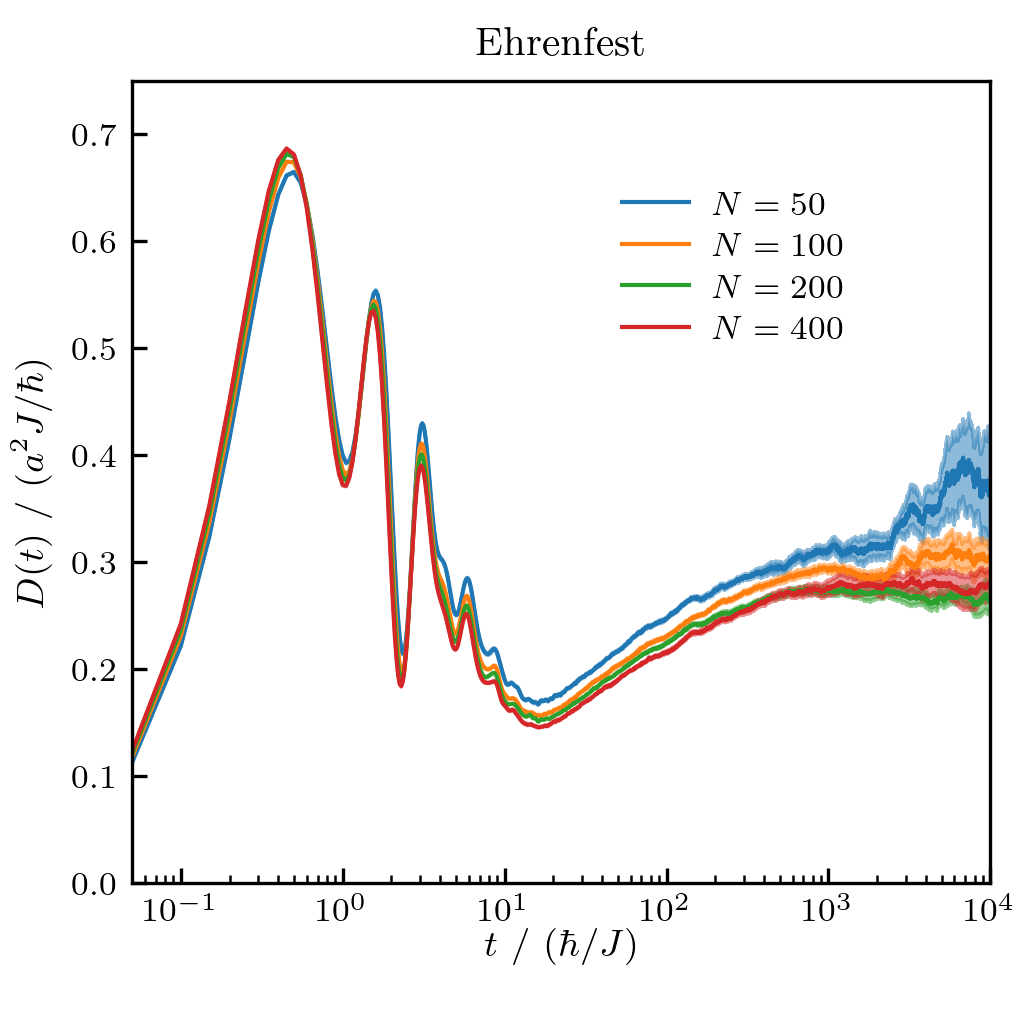}~\includegraphics[scale=0.95]{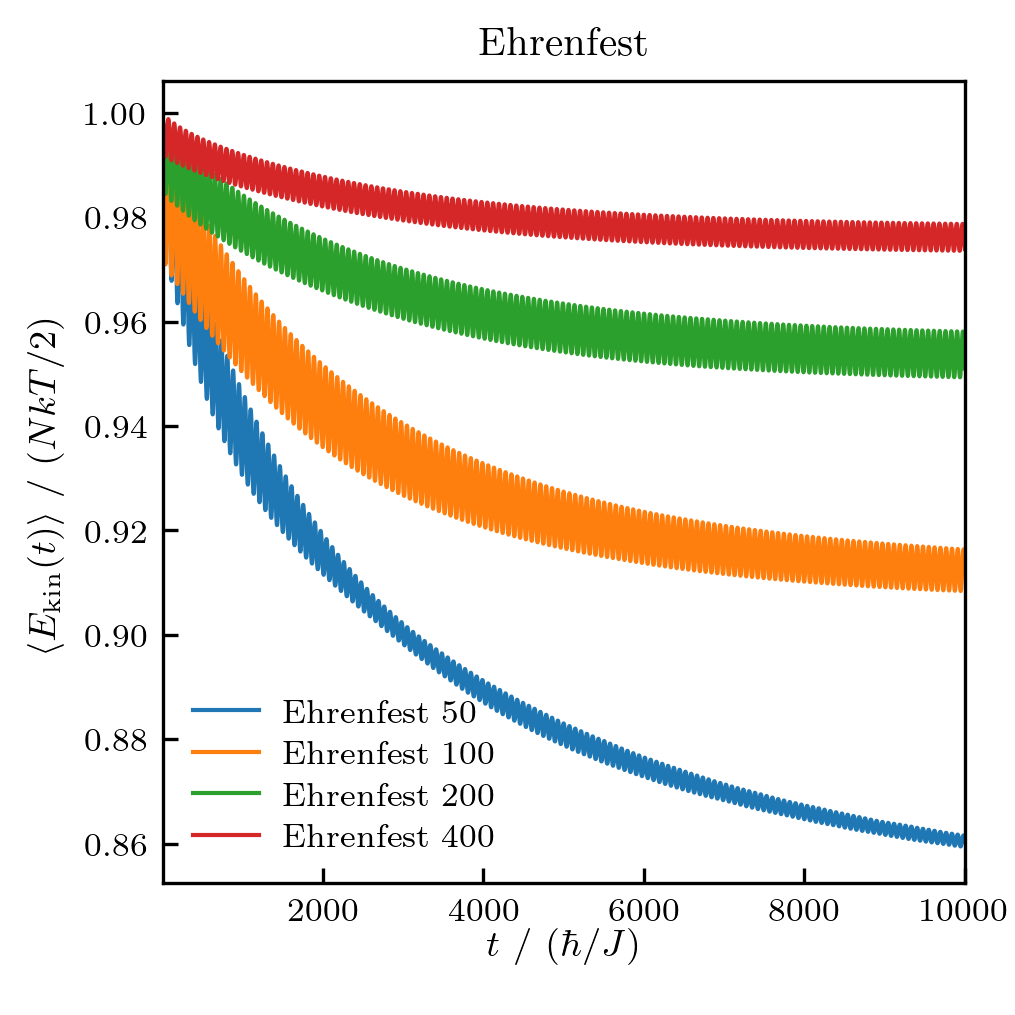}
    \caption{Left: Convergence of the Ehrenfest $D(t)$ with increasing $N$. Right: Average phonon kinetic energy along the Ehrenfest trajectories in units of $Nk_{\rm B}T/2$.}
    \label{fig:ehr}
\end{figure}

The left hand panel of Fig.~\ref{fig:ehr} shows how the Ehrenfest calculation converges with increasing $N$. The results are virtually identical for $N=200$ and $N=400$ so we used $N=200$ in Fig.~1.  The right hand panel of Fig.~\ref{fig:ehr} shows the average phonon kinetic energy as a function of time along the Ehrenfest trajectories. Energy is clearly leaking from the phonons to the electronic subsystem as the Ehrenfest calculation proceeds, consistent with the increase in $E_{\rm el}(q,c)=\langle c|\hat{V}_{\rm el}(q)|c\rangle$ shown in the inset of Fig.~1. However, since the electronic subsystem can only absorb a finite amount of energy ($\sim 2J$) from the phonons before it reaches infinite temperature ($\langle c|\hat{V}_{\rm el}(q)|c\rangle \sim 0$), the fraction of the initial phonon kinetic energy that is lost decreases with increasing $N$. In the limit as $N\to\infty$, the Ehrenfest calculation converges to a non-equilibrium steady state in which the phonons are at temperature $T$ and the electronic subsystem is at infinite temperature. 

\newpage

\subsection{Classical path approximation}

\begin{figure}[h]
    \centering
    \includegraphics[scale=0.95]{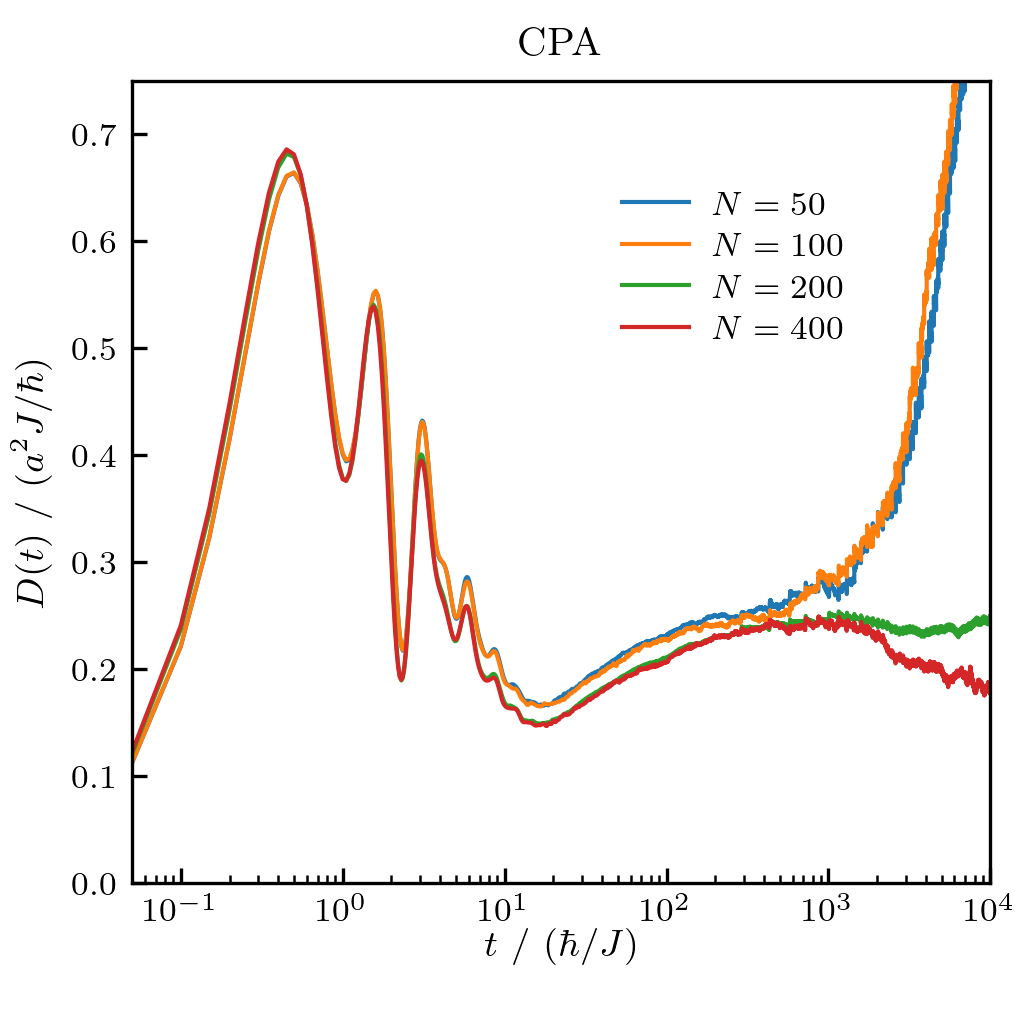}~\includegraphics[scale=0.95]{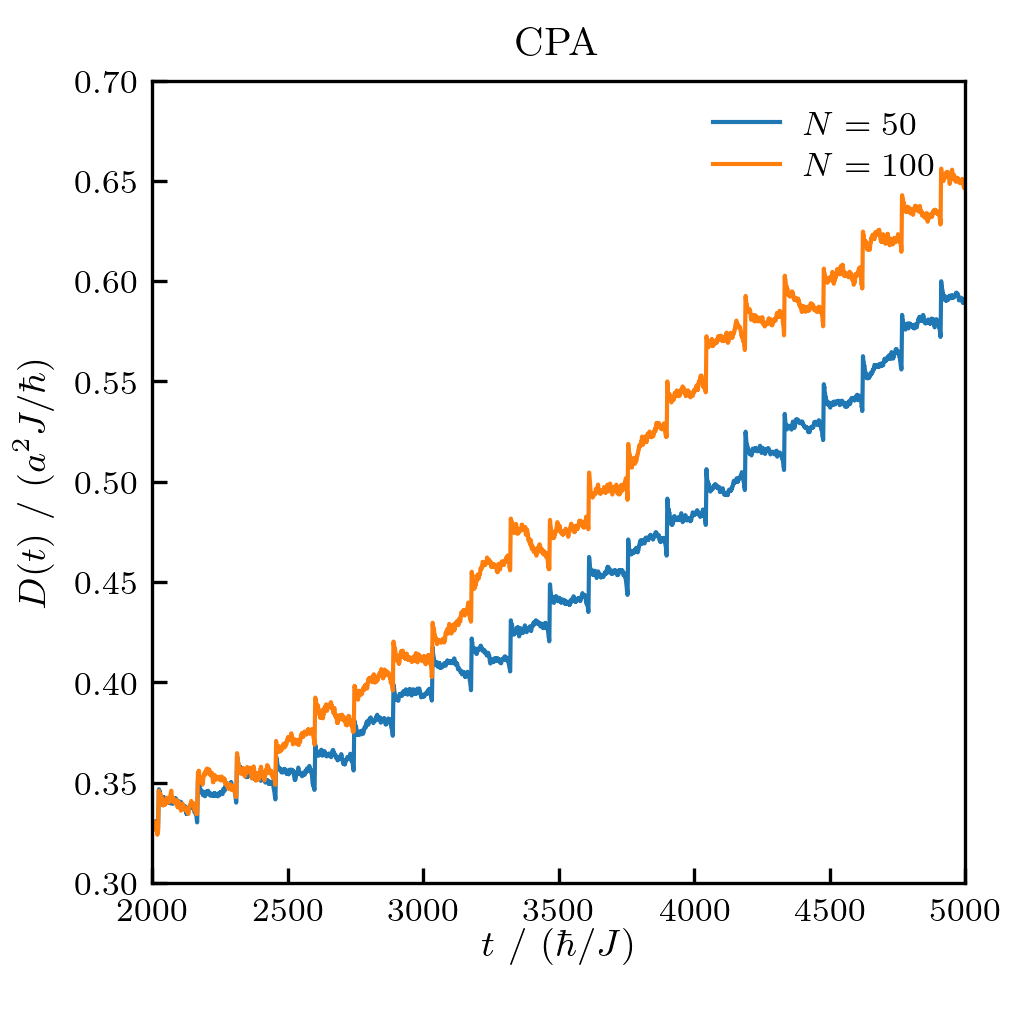}
    \caption{Left: Convergence of the CPA $D(t)$ with increasing $N$. Right: A magnified view of the finite size artefacts in the $N=50$ and $N=100$ calculations.}
    \label{fig:cpa}
\end{figure}

The left hand panel of Fig.~\ref{fig:cpa} shows how the CPA calculation converges with increasing $N$. There is a clear finite size artefact in the calculations with $N=50$ and $N=100$ which is no longer present for $N\ge 200$. For $N=400$ the long-time decrease in $D(t)$ is within the statistical errors of the calculation (not shown), so we used $N=200$ in Fig.~1. The right hand panel of Fig.~\ref{fig:cpa} zooms in on the finite size artefacts in the $N=50$ and $N=100$ calculations. These are clearly due to the pumping of the electronic subsystem by the periodically oscillating phonon field. Indeed the period of the steps in the panel is precisely $2\pi/\omega_0 = 144\,\hbar/J$. The periodic pumping is washed out when more thermally sampled oscillators are included in the chain, and it can also be removed for small $N$ by attaching a weak Langevin thermostat to the phonon dynamics (even $\gamma = 0.01\,\omega_0$, corresponding to a 1\% FWHM broadening of the phonon spectrum, suffices to remove the artefact entirely (not shown)). However, the CPA calculation again converges to a non-equilibrium steady state in which the phonons are at temperature $T$ and the electronic subsystem is at infinite temperature.

\newpage

\subsection{MASH}

\begin{figure}[h]
    \centering
    \includegraphics{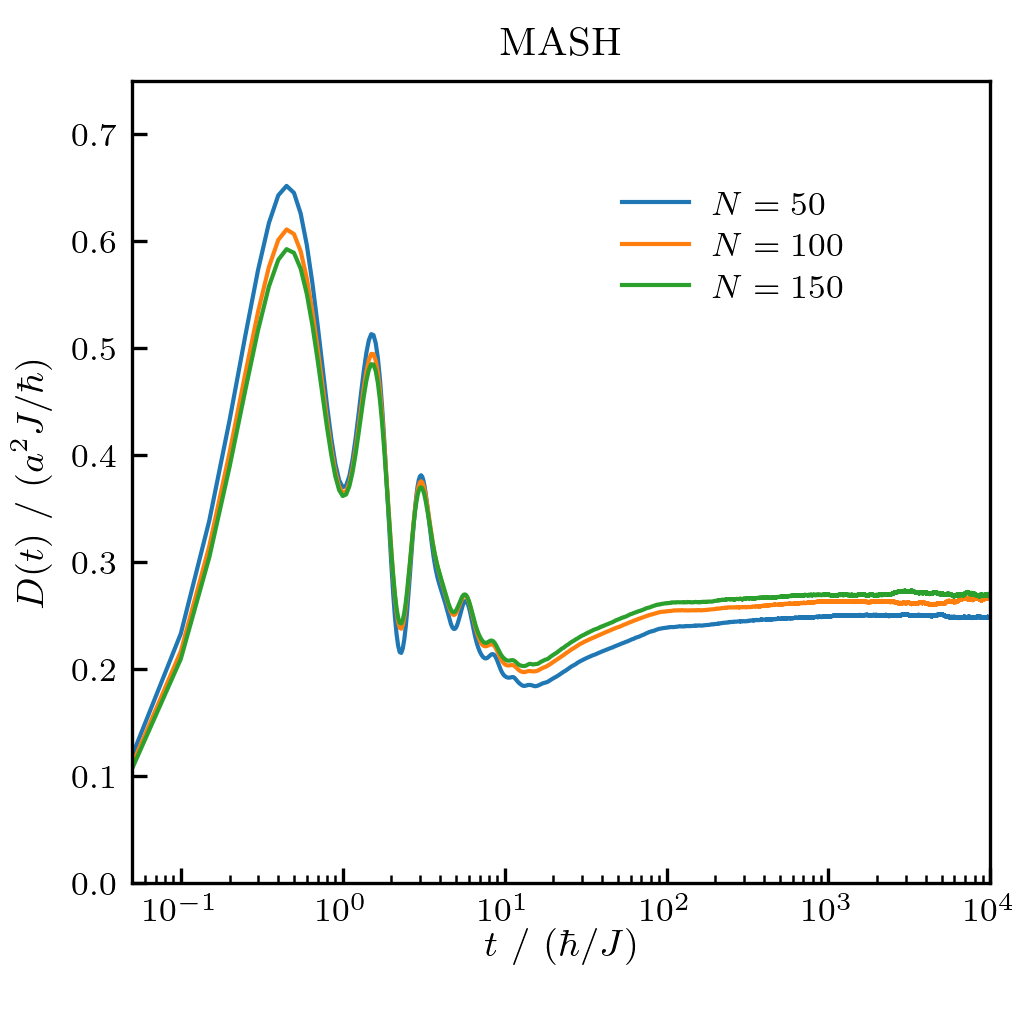}
    \caption{Convergence of the MASH $D(t)$ with increasing $N$. In this case, since the computational effort of MASH is $O(N^3)$ rather than $O(N)$, we did not go beyond $N=150$ in our convergence tests.}
    \label{fig:mash}
\end{figure}

Fig.~\ref{fig:mash} shows how the MASH calculation converges with increasing $N$. Here there are no finite size artefacts, the statistical errors are smaller, and the convergence with respect to $N$ is faster. The results are virtually identical for $N=100$ and $N=150$ so we used $N=100$ in Fig.~1 and throughout the rest of our paper. Unlike Ehrenfest dynamics or the CPA, the MASH calculation converges to a genuine quantum-classical thermal equilibrium state in which both the phonons and the electronic subsystem are at temperature $T$. 

